\documentclass[12pt]{article}
\renewcommand{\theequation}{\arabic{equation}}
\newcommand{\EQ}{\begin{equation}}
\newcommand{\EN}{\end{equation}}

\newcommand{\ket}[1]{\left|#1\right\rangle}      

\newcommand{\n}{\noindent}

\newcommand{\bear}{\begin{eqnarray}}
\newcommand{\ear}{\end{eqnarray}}
\newcommand{\bt} { \begin{tabular} }
\newcommand{\et}{ \end{tabular} }
\newcommand{\bc} { \begin{center} }
\newcommand{\ec}{ \end{center} }

\newcommand{\btb} { \begin{table} }
\newcommand{\etb}{ \end{table} }

\usepackage{graphicx}
\usepackage{graphics}
\usepackage{epsfig}

\begin{document}

\topmargin 0pt
\oddsidemargin 5mm
\newcommand{\NP}[1]{Nucl.\ Phys.\ {\bf #1}}
\newcommand{\PL}[1]{Phys.\ Lett.\ {\bf #1}}
\newcommand{\NC}[1]{Nuovo Cimento {\bf #1}}
\newcommand{\CMP}[1]{Comm.\ Math.\ Phys.\ {\bf #1}}
\newcommand{\PR}[1]{Phys.\ Rev.\ {\bf #1}}
\newcommand{\PRL}[1]{Phys.\ Rev.\ Lett.\ {\bf #1}}
\newcommand{\MPL}[1]{Mod.\ Phys.\ Lett.\ {\bf #1}}
\newcommand{\JETP}[1]{Sov.\ Phys.\ JETP {\bf #1}}
\newcommand{\TMP}[1]{Teor.\ Mat.\ Fiz.\ {\bf #1}}

\renewcommand{\thefootnote}{\fnsymbol{footnote}}

\newpage
\setcounter{page}{0}
\begin{titlepage}
\begin{flushright}
UFSCARF-TH-12-17
\end{flushright}
\vspace{0.5cm}
\begin{center}
{\large  Exact solution and finite size properties \\
of the $U_{q}[osp(2|2m)]$ vertex models}\\
\vspace{1cm}
{\large W. Galleas and M.J. Martins } \\
\vspace{1cm}
{\em Universidade Federal de S\~ao Carlos\\
Departamento de F\'{\i}sica \\
C.P. 676, 13565-905~~S\~ao Carlos-SP, Brasil}\\
\end{center}
\vspace{1.5cm}

\begin{abstract}
We have diagonalized the transfer matrix of the $U_{q}[osp(2|2m)]$ vertex model by means of 
the algebraic Bethe ansatz  method
for a variety of grading possibilities. This allowed us to investigate the
thermodynamic limit as well as the finite size properties of the corresponding
spin chain in the massless regime. The leading behaviour of the finite size
corrections to the spectrum is conjectured for arbitrary $m$. For $m=1$ we find
a critical line with central charge $c=-1$ whose exponents vary continuously with
the $q$-deformation parameter. For $m \geq 2$ the finite size term related to the conformal
anomaly depends on the anisotropy which indicates a multicritical behaviour  typical
of loop models.
\end{abstract}

\vspace{1.5cm}
\centerline{PACS numbers:  05.50+q, 02.30.IK}
\vspace{.1cm}
\centerline{Keywords: Algebraic Bethe Ansatz, Lattice Models, Finite Size}
\vspace{.15cm}
\centerline{December 2006}
\end{titlepage}

\renewcommand{\thefootnote}{\arabic{footnote}}

\section{Introduction}

Two-dimensional vertex models of statistical mechanics are nowadays considered
classical paradigms of the theory of exactly solvable models \cite{BA}.
Their statistical weights can be directly related to the elements of a $R$-matrix
satisfying the Yang-Baxter equation invariant relative to the fundamental
representations of $U_{q}[\mathcal{G}]$ quantum symmetries \cite{JI}.

The thermodynamic limit properties of most vertex models derived from ordinary
Lie algebras, such as the free-energy and the nature of the excitations,
have been well examined over the past decades in the literature, see
for instance \cite{RV,DEV,JU,KU} and references therein. It is
believed, for instance, that the massless regimes of these vertex models
are described by the critical properties of Wess-Zumino-Witten field
theories on the group $\mathcal{G}$ \cite{K7}. We remark, however, that
at least one counter example to such common belief appears to occur in the
$U_{q}[sp(2m)]$ vertex models \cite{MAR}.

By way of contrast, similar physical properties of the $U_{q}[\mathcal{G}]$
vertex models when $\mathcal{G}$ is a superalgebra have not yet been examined
in details.
The majority of the results concerning the possible universality classes of critical
behaviour governing the massless phases in these systems have been concentrated on
the $sl(n|m)$ symmetry \cite{DEV1,JU1,SA}. Similar information for other
superalgebras such as $osp(r|2m)$ has so far been restricted to the rational limit
$q \rightarrow 1$ \cite{MB,FRA,SAI}. It is not yet clear, however, if the determined
classes of universality are robust against $q$-deformations such as the cases
of ungraded algebras.

In this paper we hope to start to bridge this gap by investigating the leading finite
size corrections governing the eigenspectrum of the $U_{q}[osp(2|2m)]$
vertex models. These finite size properties have a direct relationship
with the critical operator content of massless phases \cite{CA}. In
order to do that we have diagonalized the respective row-to-row transfer
matrix by means of the algebraic Bethe ansatz approach. We have considered
explicitly all grading choices
that are compatible with the underlying
$U(1)$ symmetries of the $R$-matrix.  This step will complement our previous efforts 
concerning the Bethe ansatz solution of the $U_q[osp(n|2m)]$ vertex models \cite{GA}.
We recall that the exact solution for $n=2$ was not presented before \cite{GA} due to technical
problems with the special grading considered in that work. Here we are
able to circumvent such technicalities.

This paper is organized as follows. We start next section by describing the statistical
weights of the $U_q[osp(2|2m)]$ vertex models. In section 3 we discuss 
the
diagonalization of the corresponding row-to-row transfer matrix, within the
algebraic Bethe ansatz method, for a variety of grading possibilities.
In section 4 we use such grading freedom to choose the appropriate one in order to deal with the
thermodynamic limit in the simplest possible manner.
In section $5$ we study the finite size properties of the $U_{q}[osp(2|2m)]$ vertex models by
both analytical and numerical approaches. This provides us the basis
to conjecture, in the massless regime, the behaviour of the leading finite
size corrections to the spectrum for general $m$. For $m=1$ these results
indicate that the central charge of the underlying conformal field theory
is $c=-1$. For $m \geq 2$, however,
we find that the finite size term associated to 
the conformal central charge depends on the anisotropy coupling $q$. In Appendices A and B
we describe the technical details entering the Bethe ansatz solution of a particular
grading. 

\section{The $U_{q}[osp(2|2m)]$ vertex model}

The $R$-matrix of the $U_q[osp(2|2m)]$ vertex model is defined on the tensor
product of $Z_2$ graded spaces having two species of bosons and $2m$ species of fermions.
The Grassmann parity $p_{\alpha}$ is used to distinguish the bosonic
$p_{\alpha}=0$ and the fermionic $p_{\alpha}=1$ degrees of freedom.

To establish the statistical interpretation of this system it is  important
to know the structure of the $R$-matrix $R_{12}(\lambda)$ 
in appropriate coordinates such as the Weyl basis.
This task is in general rather involved for superalgebras but recently some
progresses towards this direction have been made \cite{GA1,LIN}. 
The Boltzmann weights of such systems can be conveniently written in terms of
the standard relation \cite{KUL},
\begin{equation}
R_{12}(\lambda)= P_{12} \check{R}_{12}(\lambda),
\end{equation}
where $P_{12}$ is the graded permutator given by
$P= \displaystyle \sum_{\alpha,\beta=1}^{N} (-1)^{p_{\alpha} p_{\beta}} \hat{e}_{\alpha\beta}
\otimes \hat{e}_{\beta\alpha}$ and 
$\hat{e}_{\alpha\beta}$
denotes $N \times N$ matrices having only one non-null
element with value 1 at row $\alpha$ and column $\beta$. The operator
$\check{R}_{12}(\lambda)$ satisfies
the following form of the Yang-Baxter equation, 
\begin{equation}
\label{ybr}
{\check{R}}_{12}(\lambda-\mu) {\check{R}}_{23}(\lambda) {\check{R}}_{12}(\mu)=
{\check{R}}_{23}(\mu) {\check{R}}_{12}(\lambda) {\check{R}}_{23}(\lambda-\mu) .
\end{equation}
which is insensitive to grading.

It turns out that the corresponding $\check{R}$-matrix of 
the $U_{q}[osp(2|2m)]$ vertex model,
in terms of the Weyl basis, can be written as,
\bear
\label{Ryb}
{\check{R}}^{(0)}(\lambda) &=&\sum_{\stackrel{\alpha=1}{\alpha \neq \alpha'}}^{N_{0}} a_{\alpha}^{(0)} (\lambda)
\hat{e}_{\alpha \alpha} \otimes \hat{e}_{\alpha \alpha}
+b^{(0)} (\lambda) \sum_{\stackrel{\alpha ,\beta=1}{\alpha \neq \beta,\alpha \neq \beta'}}^{N_{0}}
(-1)^{p_{\alpha}^{(0)} p_{\beta}^{(0)} } \hat{e}_{\beta \alpha} \otimes \hat{e}_{\alpha \beta}  \nonumber \\
&& +{\bar{c}}^{(0)} (\lambda) \sum_{\stackrel{\alpha ,\beta=1}{\alpha < \beta,\alpha \neq \beta'}}^{N_{0}} \hat{e}_{\alpha \alpha} \otimes \hat{e}_{\beta \beta}
+c^{(0)} (\lambda) \sum_{\stackrel{\alpha ,\beta=1}{\alpha > \beta,\alpha \neq \beta'}}^{N_{0}} \hat{e}_{\alpha \alpha} \otimes \hat{e}_{\beta \beta} \nonumber \\
&& + \sum_{\alpha ,\beta =1}^{N_{0}} d_{\alpha, \beta}^{(0)} (\lambda)
\hat{e}_{\alpha' \beta} \otimes \hat{e}_{\alpha \beta'} \;\; .
\ear

For later convenience we have introduced the label 
$(0)\equiv (2|2m)$. It emphasizes that we are considering a $Z_2$ graded space with two
bosonic and $2m$ fermionic degrees of freedom and $N_0=2+2m$ denotes the dimension of such space. Each
index $\alpha$ has its conjugated $\alpha' = N_0 +1 -\alpha$ and the Boltzmann weights
$a_{\alpha}^{(0)}(\lambda)$,
$b^{(0)}(\lambda)$, $c^{(0)}(\lambda)$ and ${\bar{c}}^{(0)} (\lambda)$ are given by
\begin{eqnarray}
\label{bw1}
a_{\alpha}^{(0)} (\lambda) &=&(e^{2 \lambda} -\zeta^{(0)})(e^{2 \lambda (1-p_{\alpha}^{(0)})} -q^2 e^{2\lambda p_{\alpha}^{(0)}}) \;\;\;\;\;\;\;\;\;
b^{(0)} (\lambda) = q(e^{2 \lambda} -1)(e^{2 \lambda} -\zeta^{(0)}) \nonumber \\
c^{(0)} (\lambda) &=&(1-q^2)(e^{2 \lambda} -\zeta^{(0)}) \;\;\;\;\;\;\;\;\;\;\;\;\;\;\;\;\;\;\;\;\;\;\;\;\;\;\;\;\;\;
{\bar{c}}^{(0)} (\lambda) = e^{2 \lambda} c^{(0)}(\lambda) ,
\end{eqnarray}
\n while $d_{\alpha\beta}^{(0)}(\lambda)$ has the form
\begin{equation}
\label{bw2}
d_{\alpha, \beta}^{(0)} (\lambda)=\cases{
\displaystyle  q(e^{2 \lambda} -1)(e^{2 \lambda} -\zeta^{(0)}) +e^{2\lambda}(q^2 -1)(\zeta^{(0)} -1) \;\;\;\;\;\;\;\;\;\;\;\;\;\;\;\;  \alpha=\beta=\beta' \;\; \cr
\displaystyle  (e^{2 \lambda} -1)\left[ (e^{2 \lambda} -\zeta^{(0)}) (-1)^{p_{\alpha}^{(0)}} q^{2 p_{\alpha}^{(0)}} +e^{2\lambda}(q^2 -1) \right] \;\; \;\;\;\;  \;\;  \alpha=\beta \neq \beta' \;\; \cr
\displaystyle  (q^{2 }-1)\left[ \zeta^{(0)}(e^{2 \lambda} -1)\frac{\epsilon_{\alpha}}{\epsilon_{\beta}} q^{t_{\alpha}-t_{\beta}} -\delta_{\alpha ,\beta'} (e^{2\lambda} -\zeta^{(0)}) \right] \;\;\;\;\; \; \;\;  \alpha < \beta \;\;\cr
\displaystyle  (q^{2 }-1) e^{2 \lambda} \left[ (e^{2 \lambda} -1)\frac{\epsilon_{\alpha}}{\epsilon_{\beta}} q^{t_{\alpha}-t_{\beta}} -\delta_{\alpha ,\beta'} (e^{2\lambda} -\zeta^{(0)}) \right] \;\;\; \;\;\;\;\;\;  \alpha > \beta \;\;\cr } .
\end{equation}

We stress that the formulas (\ref{Ryb}-\ref{bw2}) are valid only for 
grading choices whose 
respective parities $p_{\alpha}^{(0)}$ satisfy 
the reflexion condition $p_{\alpha}^{(0)}=p_{\alpha'}^{(0)}$. These grading possibilities
are consonant with the underlying
$U(1)$ symmetries  of the system that usually play an essential role in Bethe ansatz
solutions. Furthermore, the parameter
$\zeta^{(0)}= q^{-2m}$ and the variables
$\epsilon_{\alpha}$ and $t_{\alpha}$ are related to the parities by
\begin{eqnarray}
\epsilon_{\alpha} &=& \cases{
\displaystyle (-1)^{-\frac{p_{\alpha}^{(0)}}{2}} \;\;\; \;\;\;\;\;\;\;\;\; \;\;\;\; \;\;\;\;\;\;\;\;\;\;\; \;\;\;\;\;\;\;\;\;\;\;\;\; \;\; \;\; 1 \leq \alpha \leq \frac{N_{0}}{2} \;\; \cr
\displaystyle (-1)^{\frac{p_{\alpha}^{(0)}}{2}} \;\;\;\; \;\;\;\;\;\;\; \;\;\;\; \;\;\;\;\;\;\;\;\;\;\;\;\;\;\;\;\;\;\;\;\;\;\; \;\;\;\;\;\;\;\;  \frac{N_{0}}{2}+1 \leq \alpha \leq N_{0} \;\; \cr },
\\
t_{\alpha} &=& \cases{
\alpha + \left[ \frac{1}{2} -p_{\alpha}^{(0)} +2\displaystyle{\sum_{\alpha \leq \beta \leq \frac{N_{0}}{2}} p_{\beta}^{(0)}} \right] \;\;\;\;\; \;\;\;\;\;\; \;\;  1 \leq \alpha \leq \frac{N_{0}}{2} \;\; \cr
\alpha - \left[ \frac{1}{2} -p_{\alpha}^{(0)} +2\displaystyle{\sum_{\frac{N_{0}}{2}+1 \leq \beta \leq \alpha} p_{\beta}^{(0)}} \right] \;\;\;\;\;\;\;\; \;\;  \frac{N_{0}}{2}+1 \leq \alpha \leq N_{0} \;\; \cr } .
\label{bwf}
\end{eqnarray}

We would like to close this section with the following remark.  The above explicit expression for the
$U_q[osp(2|2m)]$ $\check{R}$-matrix was first presented by us for a particular grading choice \cite{GA}
and later on generalized to include other grading possibilities satisfying the condition
$p_{\alpha}^{(0)}=p_{\alpha'}^{(0)}$ \cite{GA1}. In the former reference we claimed also to have 
exhibited the explicit expression of the $\check{R}$-matrix associated to the twisted $U_q[osp^{(2)}(2n|2m)]$ 
quantum superalgebra. Recently, however, we realized that such identification is not correct and the
$\check{R}$-matrix denoted by $U_q[osp^{(2)}(2n|2m)]$ in \cite{GA} is in fact the one invariant relative
to 
the $U_q[spo(2n|2m)]$ quantum symmetry \footnote{
We thank J.R. Links for suggesting us that this may be the case.} \cite{SH}.   This means that the results
to be obtained in next sections are therefore also valid for 
the vertex model based on the $U_q[spo(2m|2)]$  symmetry. We believe that the correct $U_q[osp^{(2)}(2n|2m)]$
$R$-matrix were indeed obtained by us in \cite{GA1} as those associated with the generalizations of
Jimbo's $D^{(2)}_{n+1}$ $R$-matrix. We hope that this later identification could be confirmed in near
future by means a detailed analysis of the set of algebraic relations coming
from the respective Yang-Baxter algebra \cite{FRT}. We also note that
the $R$-matrices associated to $q$-deformations of the $osp(2|2)$  symmetry
have been previously investigated in \cite{DEG,LINJ}.

\section{The algebraic Bethe ansatz}

The quantum inverse scattering method provides us a systematic 
framework to construct and solve integrable vertex models
by the algebraic Bethe ansatz \cite{QIM}. It also can be extended to systems whose  
$R$-matrices are invariant relative to Lie superalgebras \cite{KUL}. In this approach
we start by considering a collection of $R$-matrices, $R_{\mathcal{A} j}(\lambda)$ with
$j=1,\dots, L$, acting non-trivially on the auxiliary space $\mathcal{A}^{(0)} \equiv \mathrm{C}^{N_{0}}$ and
on the $j$-th node of the quantum space $\displaystyle \bigotimes_{j=1}^{L} \mathrm{C}^{N_{0}}$.
An important ingredient is the  monodromy matrix defined 
by the following ordered product of $R$-matrices,
\EQ
\label{mono}
{\cal{T}}^{(0)}(\lambda)= R_{{\cal{A}}L}^{(0)}(\lambda)
R_{{\cal{A}}L-1}^{(0)}(\lambda)  \dots
R_{{\cal{A}}1}^{(0)}(\lambda). \;\; 
\EN

The 
row-to-row transfer matrix of the respective vertex model can then be written as the
supertrace  of the monodromy matrix with respect to the auxiliary space \cite{KUL}, namely 
\EQ
\label{trans}
T^{(0)}(\lambda)= \mathrm{Str}_{{\cal A}^{(0)}}[{\cal{T}}^{(0)}(\lambda)]=\sum_{\alpha =1}^{N_{0}} (-1)^{p_{\alpha}^{(0)}}
 {\cal{T}}_{\alpha \alpha}^{(0)}(\lambda).
\EN

The next step  is to present the solution of
the eigenvalue problem, 
\EQ
\label{eigenp}
T^{(0)}(\lambda) \ket{\Phi} = \Lambda^{(0)} (\lambda) \ket{\Phi},
\EN
within an algebraic formulation of the Bethe ansatz.

In this section we tackle the problem (\ref{eigenp}) 
in the case of the  
$R$-matrices (\ref{Ryb}-\ref{bwf}) of previous section for any of the  grading 
$p_{\alpha}^{(0)} =p_{\alpha'}^{(0)}$ choices. We remark that such solution for a variety of
such
gradings is in general rather intricate even for the $U_q[sl(n|m)]$ vertex model \cite{GE,GAP}.  
Here follow the nested Bethe ansatz formalism 
developed in \cite{PB} for isotropic vertex models
and recently
extended to accommodate trigonometric 
$R$-matrices based on $q$-deformed Lie superalgebras \cite{GA}. We recall, however, that
in the later reference the Bethe ansatz solution for the specific 
case of the $U_q[osp(2|2m)]$ vertex model was not presented and  here we will be
filling this gap. Considering that the main procedure has already been well explained before \cite{PB,GA}
there is no need to repeat it again in details. In what follows
we shall restrict ourselves only to the essential points 
concerning the solution of such eigenvalue problem. Fortunately, we find that the presence 
of the many grading 
possibilities 
$p_{\alpha}^{(0)} =p_{\alpha'}^{(0)}$
can still be accommodate in terms of certain recurrence relations 
envisaged by us in \cite{GA} for a specific grading choice. This relation for the  
eigenvalues of $T^{(0)}(\lambda)$ turns out to be,
\begin{eqnarray}
\label{eirec}
&&\Lambda^{(\alpha)}(\lambda , \{ \lambda_{1}^{(\alpha)},\dots, \lambda_{n_{\alpha}}^{(\alpha)} \})=
(-1)^{p^{(\alpha)}_{1}} \prod_{i=1}^{n_{\alpha}} (-1)^{p^{(\alpha)}_{1}} a_{1}^{(\alpha)} (\lambda - \lambda_{i}^{(\alpha)})
\prod_{i=1}^{n_{\alpha +1}} (-1)^{p^{(\alpha)}_{1}} \frac{a_{1}^{(\alpha)} (\lambda_{i}^{(\alpha +1)} -\lambda)}{b^{(\alpha)} (\lambda_{i}^{(\alpha +1)} -\lambda)} \nonumber \\
&&+(-1)^{p^{(\alpha)}_{N_{\alpha}}}  \prod_{i=1}^{n_{\alpha}} (-1)^{p^{(\alpha)}_{N_{\alpha}}} d_{N_{\alpha},N_{\alpha}}^{(\alpha)} (\lambda - \lambda_{i}^{(\alpha)})
\prod_{i=1}^{n_{\alpha +1}} (-1)^{p^{(\alpha)}_{N_{\alpha}}} \frac{b^{(\alpha)} (\lambda - \lambda_{i}^{(\alpha +1)})}{d_{N_{\alpha},N_{\alpha}}^{(\alpha)} (\lambda - \lambda_{i}^{(\alpha +1)})} \nonumber \\
&&+\prod_{i=1}^{n_{\alpha}} b^{(\alpha)} (\lambda - \lambda_{i}^{(\alpha)})
\prod_{i=1}^{n_{\alpha +1}}  \frac{q^{(\alpha)}}{d_{N_{\alpha},N_{\alpha}}^{(\alpha)} (\lambda - \lambda_{i}^{(\alpha +1)})}
\Lambda^{(\alpha +1)}(\lambda , \{ \lambda_{1}^{(\alpha +1)},\dots, \lambda_{n_{\alpha +1}}^{(\alpha +1)} \}) , \nonumber \\
\end{eqnarray}
while the corresponding Bethe ansatz equations  for the 
rapidities $\{ \lambda_{j}^{(\alpha)} \}$ are given by,
\begin{eqnarray}
\label{beterec}
\prod_{i=1}^{n_{\alpha-1}} (-1)^{p^{(\alpha)}_{1}} \frac{a_{1}^{(\alpha)}(\lambda_{j}^{(\alpha)} - \lambda_{i}^{(\alpha-1)})}{b^{(\alpha)}(\lambda_{j}^{(\alpha)} - \lambda_{i}^{(\alpha-1)})}=&&
\prod_{i \neq j}^{n_{\alpha }} q^{(\alpha)} (-1)^{p^{(\alpha)}_{1}+p^{(\alpha)}_{2} } \frac{a_{1}^{(\alpha+1)}(\lambda_{j}^{(\alpha)} - \lambda_{i}^{(\alpha )})}{d_{N_{\alpha},N_{\alpha}}^{(\alpha)}(\lambda_{j}^{(\alpha)} - \lambda_{i}^{(\alpha)})}
\frac{b^{(\alpha )}(\lambda_{i}^{(\alpha)} - \lambda_{j}^{(\alpha )})}{a_{1}^{(\alpha)}(\lambda_{i}^{(\alpha)} - \lambda_{j}^{(\alpha)})} \nonumber \\
&& \times \prod_{i=1}^{n_{\alpha+1}} (-1)^{p^{(\alpha+1)}_{1}} \frac{a_{1}^{(\alpha+1)}(\lambda_{i}^{(\alpha+1)} - \lambda_{j}^{(\alpha+1)})}{b^{(\alpha+1)}(\lambda_{i}^{(\alpha+1)} - \lambda_{j}^{(\alpha+1)})} \;\; .
\end{eqnarray}

We now  describe the way the recurrence relations (\ref{eirec},\ref{beterec}) should be interpreted. The label
$(\alpha)$ in the eigenvalues and Bethe ansatz equations was introduced to characterize the respective graded
space these results are concerned with. The dimension of such space is twice less than the one we started with,
$N_{\alpha}=N_{0} - 2\alpha$, and the respective number of bosonic and fermionic degrees of freedom are determined
by the following rule,
\begin{equation}
\label{lala}
(\alpha) \equiv
\displaystyle ( N_{\alpha} - \sum_{\beta=1}^{N_{\alpha}} p_{\beta}^{(\alpha)} |
\sum_{\beta=1}^{N_{\alpha}} p_{\beta}^{(\alpha)} ).
\end{equation}

The Grassmann parities 
$p_{\beta}^{(\alpha)}$ associated with the
graded space $(\alpha)$ are obtained through the relation $p_{\beta}^{(\alpha+1)}= p_{\beta+1}^{(\alpha)}$ for
$\beta=1, \dots, N_{\alpha}-2$. 
The Boltzmann weights $a_{1}^{(\alpha)}(\lambda)$, $b^{(\alpha)}(\lambda)$ and
$d_{N_{\alpha},N_{\alpha}}^{(\alpha)}(\lambda)$ are derived from 
(\ref{Ryb}-\ref{bwf}), considering the graded space characterized by $(\alpha)$ instead of the
original one labeled $(0)$. Finally   
$q^{(\alpha)}=(-1)^{p_1^{(\alpha)}} q^{1-2p_1^{(\alpha)}}$ and the consistency with the
original eigenvalue problem requires us to set 
$\lambda_j^{(0)} =0$ for $j=1, \dots, n_{0}$ and to make the 
identification $n_{0} \equiv L$.

In order to obtain the eigenvalues and respective Bethe ansatz equations for a given choice
of parities 
$p_{\alpha}^{(0)} =p_{\alpha'}^{(0)}$ we need to iterate the relations (\ref{eirec},\ref{beterec}) starting
from $\alpha=0$. We then  carry on such nested procedure
until we reach a final step labeled by $(f)$ and therefore up to
$\alpha=f-1$. In this last step we have to deal with the diagonalization of  an
inhomogeneous transfer matrix of the following type,
\begin{eqnarray}
\label{tfinal}
&& T^{(f)}(\lambda,\{\lambda_1^{(f)}, \dots, \lambda_{n_{f}}^{(f)} \}) \nonumber \\
&& =Str_{{\mathcal A}^{(f)}} \left[ R^{(f)}_{{\mathcal A}^{(f)} n_{f}} (\lambda-\lambda^{(f)}_{n_{f}})  R^{(f)}_{{\mathcal A}^{(f)} n_{f}-1} (\lambda-\lambda^{(f)}_{n_{f}-1})
\dots R^{(f)}_{{\mathcal A}^{(f)} 1} (\lambda-\lambda^{(f)}_{1}) \right] .
\end{eqnarray}

The solution  of the eigenvalue problem for such last step depends much on the choice of the parities we
started with. We find that for all gradings choices satisfying 
$p_{\alpha}^{(0)} = p_{\alpha'}^{(0)}$, except the special case 
$p_{\alpha}^{(0)} = 1~~\mathrm{for}~~\alpha=1,\dots,m,m+3,\dots,2m+2$ and $p_{m+1}^{(0)}= 
p_{m+2}^{(0)}= 0$, the last step consists in the diagonalization of a common six-vertex model. 
In our notation it is identified as 
$(f) \equiv (0|2)$ and the $R$-matrix governing such
final step is,
\begin{equation}
R^{(f)}(\lambda)=\pmatrix{
a_{1}^{(f)}(\lambda) & 0 & 0 & 0  \cr
0 & d_{1,1}^{(f)}(\lambda) & d_{1,2}^{(f)}(\lambda) & 0  \cr
0 & d_{2,1}^{(f)}(\lambda) & d_{1,1}^{(f)}(\lambda) & 0  \cr
0 & 0 & 0 & a_{1}^{(f)}(\lambda) \cr},
\end{equation} 
\n with the following Boltzmann weights
\begin{eqnarray}
a_{1}^{(f)}(\lambda) &=& (e^{2 \lambda} - q^{-4}) (e^{2 \lambda} q^2 -1)  \;\;\;\;\;\;\;
d_{1,1}^{(f)}(\lambda) = \frac{1}{q^2} (e^{2 \lambda} - 1)(e^{2 \lambda} q^2 -1) \nonumber \\
d_{1,2}^{(f)}(\lambda) &=& \frac{1}{q^4} (q^{4} - 1)(e^{2 \lambda} q^2 -1) \;\;\;\;\;\;\;\;
d_{2,1}^{(f)}(\lambda) = \frac{1}{q^4} e^{2 \lambda} (q^{4} - 1)(e^{2 \lambda} q^2 -1) .
\end{eqnarray}

Considering that the Bethe ansatz solution of the six vertex model has been already well examined in the
literature, we shall not extend over this problem. 
In order to present our results in a more suitable form we define  
$\displaystyle Q_{\alpha}(\lambda)=\prod_{i=1}^{n_{\alpha}} \sinh{(\lambda -\lambda_{i}^{(\alpha)})}$
and set $q=e^{i\gamma}$. In this way we have the following expression for the eigenvalues
\begin{eqnarray}
\label{grad1}
\Lambda^{(0)} (\lambda) &=& (-1)^{p^{(0)}_{1}} \left[ (-1)^{p^{(0)}_{1}} a_{1}^{(0)}(\lambda) \right]^{L}
\frac{Q_{1}\left( \lambda + (-1)^{p^{(0)}_{1}} i\frac{\gamma}{2} \right)}{Q_{1}\left( \lambda - (-1)^{p^{(0)}_{1}} i\frac{\gamma}{2} \right)} \nonumber \\
&+& (-1)^{p^{(0)}_{N_{0}}} \left[ (-1)^{p^{(0)}_{N_{0}}} d_{N_{0},N_{0}}^{(0)}(\lambda) \right]^{L}
\frac{Q_{1}\left( \lambda + (2m - (-1)^{p^{(0)}_{1}}) i \frac{\gamma}{2} \right)}{Q_{1}\left( \lambda + (2m + (-1)^{p^{(0)}_{1}}) i \frac{\gamma}{2} \right) } \nonumber \\
&+& \left[ b^{(0)} (\lambda)  \right]^{L} \sum_{\alpha=1}^{2m} G_{\alpha}(\lambda | \{ \lambda_{j}^{(\beta)} \}) \nonumber \\
\end{eqnarray}
where the auxiliary functions
$G_{\alpha}(\lambda | \{ \lambda_{j}^{(\beta)} \}) $ are given by,
\begin{eqnarray}
&& G_{\alpha}(\lambda | \{ \lambda_{j}^{(\beta)} \}) \nonumber \\
&& =\cases{
(-1)^{p_{\alpha+1}^{(0)}} \frac{Q_{\alpha} (\lambda - \delta_{\alpha} - (-1)^{p_{\alpha+1}^{(0)}} i \gamma )}{Q_{\alpha} \left(\lambda -\delta_{\alpha}  \right)}
\frac{Q_{\alpha+1} (\lambda  - \delta_{\alpha+1} + (-1)^{p_{\alpha+1}^{(0)}} i \gamma  )}{Q_{\alpha+1} \left(\lambda - \delta_{\alpha+1}  \right)}
\;\;\;\;\;\;\;\;\;\;\;\;\;\;\;\;\;\;\;\;\;\;\;\; \alpha = 1,\dots ,m-1 \cr
-\frac{Q_{m} \left(\lambda -\delta_{m} + i\gamma  \right)}{Q_{m} \left(\lambda - \delta_{m}  \right)}
\frac{Q_{m+1} \left(\lambda - \delta_{m+1} - 2 i \gamma  \right)}{Q_{m+1} \left(\lambda - \delta_{m+1}  \right)}
\;\;\;\;\;\;\;\;\;\;\;\;\;\;\;\;\;\;\;\;\;\;\;\;\;\;\;\;\;\;\;\;\;\;\;\;\;\;\;\;\;\;\;\;\;\;\;\;\;\;\;\;\; \alpha = m  \cr
G_{\alpha - m}(-i m \gamma -\lambda |- \{ \lambda_{j}^{(\beta)} \}) \;\;\;\;\;\;\;\;\;\;\;\;\;\;\;\;\;\;\;\;\;\;\;\;\;\;\;\;\;\;\;\;\;\;\;\;\;\;\;\;\;\;\;\;\;\;\;\;\;\;\;\;\;\;\;\;\; \alpha = m+1,\dots ,2m \cr } \nonumber
\end{eqnarray}

The rapidities $\left\{ \lambda_{j}^{(\alpha)} \right\}$ are constrained to satisfy the following
set of Bethe ansatz equations,
\begin{eqnarray}
\label{bffb}
\prod_{i=1}^{n_{\alpha-1}} \frac{\sinh{\left(\lambda_{j}^{(\alpha)} -\lambda_{i}^{(\alpha-1)} - (-1)^{p_{\alpha}^{(0)}} i\frac{\gamma}{2} \right)}}{\sinh{\left(\lambda_{j}^{(\alpha)} -\lambda_{i}^{(\alpha-1)} + (-1)^{p_{\alpha}^{(0)}} i\frac{\gamma}{2} \right)}} &=&
\prod_{i\neq j}^{n_{\alpha}} \frac{\sinh{\left(\lambda_{j}^{(\alpha)} -\lambda_{i}^{(\alpha)} +i k_{\alpha} \gamma \right)}}{\sinh{\left(\lambda_{j}^{(\alpha)} -\lambda_{i}^{(\alpha)} -i k_{\alpha} \gamma \right)}} \nonumber \\
&\times& \prod_{i=1}^{n_{\alpha+1}} \frac{\sinh{\left(\lambda_{i}^{(\alpha+1)} -\lambda_{j}^{(\alpha)} - (-1)^{p_{\alpha+1}^{(0)}} i g_{\alpha}\frac{\gamma}{2} \right)}}{\sinh{\left(\lambda_{i}^{(\alpha+1)} -\lambda_{j}^{(\alpha)} + (-1)^{p_{\alpha+1}^{(0)}} i g_{\alpha} \frac{\gamma}{2} \right)}} \nonumber \\
&& \;\;\;\;\;\;\;\;\;\;\;\;\;\;\;\;\;\;\;\;\;\;\;\;\;\;\;\;\;\;\;\;\;\;\;\;\;\;\;\;\;\;\;\;\;\;\;\;\;\;\;\;\; \alpha= 1,\dots , m  \nonumber \\
\prod_{i=1}^{n_{m}} \frac{\sinh{\left(\lambda_{j}^{(m+1)} - \lambda_{i}^{(m)} + i\gamma \right)}}{\sinh{\left(\lambda_{j}^{(m+1)} - \lambda_{i}^{(m)} - i\gamma \right)}}&=&
\prod_{i \neq j}^{n_{m+1}} \frac{\sinh{\left( \lambda_{j}^{(m+1)} - \lambda_{i}^{(m+1)} + 2i\gamma  \right)}}{\sinh{\left( \lambda_{j}^{(m+1)} - \lambda_{i}^{(m+1)} - 2i\gamma     \right)}}
\end{eqnarray}
where 
$k_{\alpha}= -\frac{1}{2} \left[ (-1)^{p_{\alpha}^{(0)}} + (-1)^{p_{\alpha+1}^{(0)}} \right]$ and
$g_{\alpha}=\cases{
2 \;\;\;\;\;\;\; \alpha=m \cr
1 \;\;\;\;\;\;\; \mbox{otherwise}\cr}$.

We remark that in order to obtain the Bethe ansatz equation in the above symmetric form we
have performed the 
shifts $\{ \lambda_j^{(\alpha)} \} \rightarrow
\{ \lambda_j^{(\alpha)} \}+\delta_{\alpha}$. The variables 
$\delta_{\alpha}$ have a strong dependence on the parities and are given by,
\begin{equation}
\delta_{\alpha}=\cases{
\displaystyle i\frac{\gamma}{2} \sum_{\beta=1}^{\alpha} (-1)^{p_{\beta}^{(0)}} \;\;\;\;\;\;\;\;\;\;\;\;\;\;\;\;\;\;\;\;\;\; \alpha=1,\dots , m \cr
\displaystyle i\frac{\gamma}{2} \left[ \sum_{\beta=1}^{m} (-1)^{p_{\beta}^{(0)}} - 2 \right] \;\;\;\;\;\;\;\;\;\;\;\; \alpha=m+1 \cr}.
\end{equation}

As usual we see that the Bethe ansatz equations as well as the eigenvalues depend strongly
on choice of the parities 
$p_{\alpha}^{(0)}$. This feature has been captured here in a unified way by the index $k_{\alpha}$.
The possible different forms of Bethe ansatz equations concerning 
the distinct grading choices can be better appreciated in terms of Dynkin diagrams. In this representation
the
scattering factors between the rapidities $\lambda_{i}^{(\alpha)}$ 
and $\lambda_{j}^{(\beta)}$ 
are recasted in terms of the elements 
$\hat{e}_{\alpha \beta} $ of the respective Cartan matrix. In order to be more specific we exhibit in
Figure 1 the diagram related to the grading
\begin{equation}
p_{\alpha}^{(0)}=\cases{
0 \;\;\;\;\;\;\;\; \alpha=1, \; N_{0} \cr
1 \;\;\;\;\;\;\;\; \mbox{otherwise} \cr}
\end{equation}

\begin{figure}[ht]
\begin{center}
\includegraphics{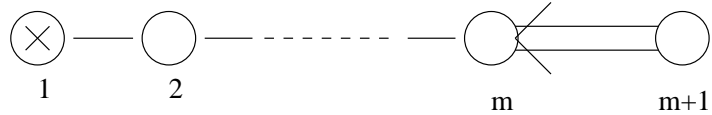}
\end{center}
\caption{\footnotesize{Representation of the Bethe ansatz equations (\ref{bffb})  in the grading $B F\dots F F \dots B $.}}
\end{figure}

The other grading possibilities in the family considered so far, namely
\begin{equation}
p_{\alpha}^{(0)}=\cases{
0 \;\;\;\;\;\;\;\; \alpha=\beta ,\beta ' \;\;\;\;\;\; \beta > 1 \cr
1 \;\;\;\;\;\;\;\; \mbox{otherwise} \cr},
\end{equation}
are represented in Figure 2.

\begin{figure}[ht]
\begin{center}
\includegraphics{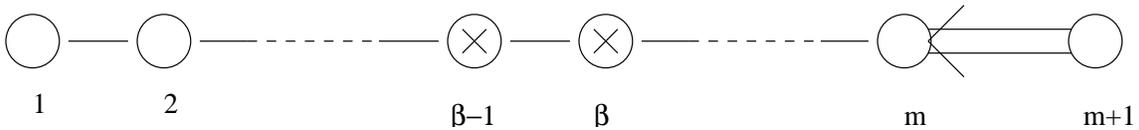}
\end{center}
\caption{\footnotesize{Representation of Bethe ansatz equations (\ref{bffb}) in the grading $F\dots F B F \dots F F \dots F B F \dots F$.}}
\end{figure}

We now turn our attention to the Bethe ansatz solution for the remaining grading,
\begin{equation}
\label{gd}
p_{\alpha}^{(0)}=\cases{
0 \;\;\;\;\;\;\;\; \alpha=m+1,m+2 \cr
1 \;\;\;\;\;\;\;\; \mbox{otherwise} \cr}.
\end{equation}

For the grading choice (\ref{gd}) the last step is no longer governed by the
six-vertex model. In this last stage one has to deal with a $16 \times 16$
$R$-matrix $R^{(f)}(\lambda)$ whose graded space is $(f) \equiv (2|2)$. This problem 
is in fact a special case of the one associated with the general $U_q[osp(2|2)]$  $R$-matrix
built from the admissible one-parameter four-dimensional representation \cite{LINJ}. 
The Bethe ansatz solution of such vertex model in the grading (\ref{gd}) involves extra
technicalities such as the presence of auxiliary transfer matrices that cannot be written
as trace of monodromy operators. Here we avoid overcrowding this section
with more technical details and we summarized them in Appendices A and B. In order 
to solve
the nested problem for the $U_q[osp(2|2m)]$ vertex model 
one needs to use the final
results given in Eqs.(\ref{VAI1},\ref{VAI2}) together with the recurrence relations 
(\ref{eirec},\ref{beterec}). By performing
these steps we find that the corresponding eigenvalues in the grading (\ref{gd}) are,
\begin{eqnarray}
\label{res}
\Lambda^{(0)} (\lambda) &=& \Lambda_{0} (\lambda) - \left[ b^{(0)}(\lambda) \right]^L \sum_{\alpha=1}^{2m} G_{\alpha}(\lambda | \{ \lambda_{j}^{(\beta)} \}) \nonumber
\end{eqnarray}
\begin{eqnarray}
\Lambda_{0} (\lambda) = \cases{
- \left[ - a_{1}^{(0)}(\lambda) \right]^{L} \frac{Q_{1}\left( \lambda - i\frac{\gamma}{2}\right)}{Q_{1}\left( \lambda + i\frac{\gamma}{2}\right)}
- \left[ - d_{N_{0},N_{0}}^{(0)}(\lambda) \right]^{L} \frac{Q_{1}\left( \lambda +(2 m +1) i\frac{\gamma}{2}\right)}{Q_{1}\left( \lambda + (2 m -1) i\frac{\gamma}{2}\right)}
\;\;\;\;\;\;\;\;\;\;\;\;\;\;\;\;\;\;\;\;\;\;\;\;\;\;\; m > 1 \cr
- \left[ - a_{1}^{(0)}(\lambda) \right]^{L} \frac{Q_{+}\left( \lambda - i\frac{\gamma}{2}\right)}{Q_{+}\left( \lambda + i\frac{\gamma}{2}\right)}
\frac{Q_{-}\left( \lambda - i\frac{\gamma}{2}\right)}{Q_{-}\left( \lambda + i\frac{\gamma}{2}\right)}
- \left[ - d_{N_{0},N_{0}}^{(0)}(\lambda) \right]^{L}  \frac{Q_{+}\left( \lambda + i\frac{3}{2} \gamma \right)}{Q_{+}\left( \lambda + i\frac{\gamma}{2} \right)}
\frac{Q_{-}\left( \lambda + i\frac{3}{2} \gamma \right)}{Q_{-}\left( \lambda + i\frac{\gamma}{2} \right)}
\;\;\;\;\; m = 1 \cr} \nonumber \\
\end{eqnarray}
\begin{eqnarray}
&& G_{\alpha}(\lambda | \{ \lambda_{j}^{(\beta)} \}) \nonumber \\
&& =\cases{
\frac{Q_{\alpha} \left(\lambda + (\alpha +2)i\frac{\gamma}{2} \right)}{Q_{\alpha} \left(\lambda + \alpha i \frac{\gamma}{2}  \right)}
\frac{Q_{\alpha+1} \left(\lambda  +(\alpha-1)i \frac{\gamma}{2} \right)}{Q_{\alpha+1} \left(\lambda + (\alpha+1)i \frac{\gamma}{2}  \right)}
\;\;\;\;\;\;\;\;\;\;\;\;\;\;\;\;\;\;\;\;\;\;\;\;\;\;\;\;\;\;\;\;\;\;\;\;\;\;\;\;\;\;\;\;\;\;\;\;\;\;\;  \alpha = 1,\dots ,m-2 \cr
\frac{Q_{m-1} \left(\lambda  + (m+1)i\frac{\gamma}{2} \right)}{Q_{m-1} \left(\lambda + (m-1)i \frac{\gamma}{2}  \right)}
\frac{Q_{+} \left(\lambda + (m-2)i\frac{\gamma}{2} \right)}{Q_{+} \left(\lambda + m i \frac{\gamma}{2} \right)}
\frac{Q_{-} \left(\lambda + (m-2)i\frac{\gamma}{2} \right)}{Q_{-} \left(\lambda + m i \frac{\gamma}{2} \right)}
\;\;\;\;\;\;\;\;\;\;\;\;\;\;\;\;\;\;\;\;\;\;\;\;\;\;\;\;  \alpha = m-1  \cr
-\frac{Q_{+} \left(\lambda + (m-2)i\frac{\gamma}{2} \right)}{Q_{+} \left(\lambda + m i \frac{\gamma}{2} \right)}
\frac{Q_{-} \left(\lambda + (m+2)i\frac{\gamma}{2} \right)}{Q_{-} \left(\lambda + m i \frac{\gamma}{2} \right)}
\;\;\;\;\;\;\;\;\;\;\;\;\;\;\;\;\;\;\;\;\;\;\;\;\;\;\;\;\;\;\;\;\;\;\;\;\;\;\;\;\;\;\;\;\;\;\;\;\;  \alpha = m  \cr
G_{\alpha - m}(-i m \gamma -\lambda |- \{ \lambda_{j}^{(\beta)} \}) \;\;\;\;\;\;\;\;\;\;\;\;\;\;\;\;\;\;\;\;\;\;\;\;\;\;\;\;\;\;\;\;\;\;\;\;\;\;\;\;\;\;\;\;\;\;\;\;\;\;\;\;  \alpha = m+1,\dots ,2m \cr }, \nonumber
\end{eqnarray}
provided the rapidities $\{ \lambda_j^{(\alpha)} \}$ satisfy the following
Bethe ansatz equations,
\begin{eqnarray}
\label{fbbf}
\prod_{i=1}^{n_{\alpha -1}} \frac{\sinh{\left(\lambda_{j}^{(\alpha)} -\lambda_{i}^{(\alpha -1)} + i\frac{\gamma}{2} \right)}}{\sinh{\left(\lambda_{j}^{(\alpha)} -\lambda_{i}^{(\alpha -1)} - i\frac{\gamma}{2} \right)}} &=&
\prod_{i\neq j}^{n_{\alpha}} \frac{\sinh{\left(\lambda_{j}^{(\alpha)} -\lambda_{i}^{(\alpha)} + i \gamma \right)}}{\sinh{\left(\lambda_{j}^{(\alpha)} -\lambda_{i}^{(\alpha)} - i \gamma \right)}} \nonumber \\
&\times& \prod_{i=1}^{n_{\alpha +1}} \frac{\sinh{\left(\lambda_{i}^{(\alpha+1)} -\lambda_{j}^{(\alpha)} + i \frac{\gamma}{2} \right)}}{\sinh{\left(\lambda_{i}^{(\alpha+1)} -\lambda_{j}^{(\alpha)} -i \frac{\gamma}{2} \right)}} 
\;\;\;\;\;\;\;\;\;\;\;\;\;\; \alpha= 1,\dots , m -2 \nonumber \\
\prod_{i=1}^{n_{m-2}} \frac{\sinh{\left(\lambda_{j}^{(m-2)} - \lambda_{i}^{(m-1)} + i\frac{\gamma}{2} \right)}}{\sinh{\left(\lambda_{j}^{(m-2)} - \lambda_{i}^{(m-1)} - i\frac{\gamma}{2} \right)}}&=&
\prod_{i \neq j}^{n_{m-1}} \frac{\sinh{\left( \lambda_{j}^{(m-1)} - \lambda_{i}^{(m-1)} + i\gamma  \right)}}{\sinh{\left( \lambda_{j}^{(m-1)} - \lambda_{i}^{(m-1)} - i\gamma   \right)}} \nonumber \\
&\times& \prod_{i=1}^{n_{+}} \frac{\sinh{\left(\lambda_{i}^{(+)} -\lambda_{j}^{(m-1)} + i \frac{\gamma}{2} \right)}}{\sinh{\left(\lambda_{i}^{(+)} -\lambda_{j}^{(m-1)} -i \frac{\gamma}{2} \right)}}
\prod_{i=1}^{n_{-}} \frac{\sinh{\left(\lambda_{i}^{(-)} -\lambda_{j}^{(m-1)} + i \frac{\gamma}{2} \right)}}{\sinh{\left(\lambda_{i}^{(-)} -\lambda_{j}^{(m-1)} -i \frac{\gamma}{2} \right)}} \nonumber \\
\prod_{i=1}^{n_{m-1}} \frac{\sinh{\left(\lambda_{j}^{(\pm)} - \lambda_{i}^{(m-1)} + i\frac{\gamma}{2} \right)}}{\sinh{\left(\lambda_{j}^{(\pm)} - \lambda_{i}^{(m-1)} - i\frac{\gamma}{2} \right)}}&=&
\prod_{i=1}^{n_{\mp}} \frac{\sinh{\left(\lambda_{i}^{(\mp)} -\lambda_{j}^{(\pm)} - i \gamma \right)}}{\sinh{\left(\lambda_{i}^{(\mp)} -\lambda_{j}^{(\pm)} + i \gamma \right)}}
\end{eqnarray}

We recall that the symmetrical form (\ref{fbbf}) is obtained after performing the 
shifts $\{ \lambda_j^{(\alpha)} \} \rightarrow
\{ \lambda_j^{(\alpha)} \} - i \alpha \frac{\gamma}{2}$ for $\alpha= 1,\dots, m-1$, and
$\{ \lambda_j^{(\pm)} \} \rightarrow
\{ \lambda_j^{(\pm)} \} - i m \frac{\gamma}{2}$. We close this section by presenting in Figure 3 the
diagrammatic representation of the Bethe ansatz equations (\ref{fbbf}).

\begin{figure}[ht]
\begin{center}
\includegraphics{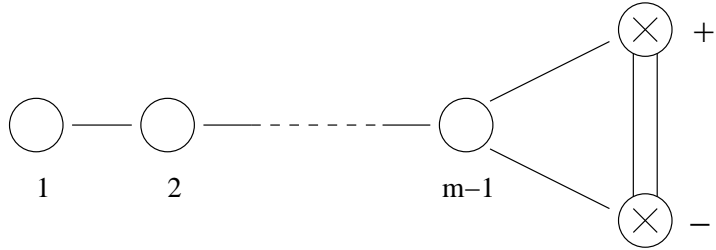}
\end{center}
\caption{\footnotesize{Representation of Bethe ansatz equations (\ref{fbbf}) in the grading $F\dots F B B F \dots F$.}}
\end{figure}

\section{Thermodynamic Limit}

In this section we will study the thermodynamic limit properties of 
the spin chain associated to the $U_q[osp(2|2m)]$ vertex model. The
corresponding Hamiltonian is formally obtained as  
the logarithmic derivative of the
transfer matrix (\ref{trans}) at the regular point $\lambda=0$,
\begin{equation}
\label{ham}
\mathcal{H} = -J [T^{(0)}]^{-1} \frac{d}{d\lambda}T^{(0)}(\lambda) \mid_{\lambda=0},
\end{equation}
where from now on we have fixed the normalization $J=\sin(\gamma)$. 

We start our analysis by studying the spectrum of 
the operator (\ref{ham}) for small
chains, 
in  the anti-ferromagnetic regime $0 \leq \gamma \leq \frac{\pi}{2}$, by means of
exact diagonalization methods for $m=1,2$. The next step is to reproduce 
the lowest energies
within the Bethe ansatz solutions  of previous section in order to
find the pattern of the corresponding roots $\{ \lambda_j^{(\alpha)} \}$.  
This helps us to select,
among the possible forms of the Bethe ansatz equations, the one
that  has  
the less complicated root structure as possible. 
This study leads us to
select the set of Bethe ansatz solution associated to the grading
$F\dots F B B F \dots F$, since the low-lying spectrum of
$(\ref{ham})$ is reproduced by using mainly real roots for all
nested levels. In this case, the eigenvalues $E^{(m)}(L,\gamma)$ of
the Hamiltonian (\ref{ham}), up to an additive 
constant, are given in terms of
the variables $\{ \lambda_j^{(\alpha)} \} $ by,
\begin{eqnarray}
\label{ene}
E^{(m)}(L,\gamma)=\cases{
\displaystyle  \sum_{i=1}^{n_{1}} \epsilon(\lambda_{i}^{(1)},\gamma)  \;\;\;\;\;\;\;\;\;\;\;\;\;\;\;\;\;\;\;\;\;\;\;\;\;\;\;\;\;\;\;\; m >1    \cr
\displaystyle  \sum_{i=1}^{n_{+}} \epsilon(\lambda_{i}^{(+)},\gamma)
+ \sum_{i=1}^{n_{-}} \epsilon(\lambda_{i}^{(-)},\gamma) \;\;\;\;\;\;\;\; m=1 \cr}
\end{eqnarray}
where $\epsilon (\lambda,\gamma)=\frac{-2J\sin{\left(\gamma \right)}}{\cosh{\left(2 \lambda \right)} - \cos{\left( \gamma \right)}}$.

We now explore the Bethe ansatz equations on the grading 
$F\dots F B B F \dots F$ in order to determine analytically 
the ground state energy and the nature of the low-energy excitations. 
Considering that the low-lying spectrum is described mostly in terms
of real roots we take directly the 
logarithmic 
of the original Bethe ansatz equations (\ref{fbbf}) and as result we find,
\begin{eqnarray}
\delta_{l,1} L \Phi(\lambda_{j}^{(l)}, \frac{\gamma}{2})=2 \pi Q_{j}^{(l)}
+\sum_{k=1}^{n_{l}} \Phi (\lambda_{j}^{(l)} - \lambda_{k}^{(l)},\gamma) 
-\sum_{\alpha=l\pm 1} \sum_{k=1}^{n_{\alpha}} \Phi (\lambda_{j}^{(l)} - \lambda_{k}^{(\alpha)},\frac{\gamma}{2}) \nonumber \\
\;\;\;\;\;\;\;\;\;\;\;\;\;\;\;\;\;\;\;\;\;\;\;\;\;\;\;\;\;\;\;\;\;\;\;\;\;\;\;\;\;\;\;\;l=1,\dots,m-2 
\label{be1}
\end{eqnarray}
\begin{eqnarray}
\delta_{l,1} L \Phi(\lambda_{j}^{(l)}, \frac{\gamma}{2})=2 \pi Q_{j}^{(l)}
+\sum_{k=1}^{n_{l}} \Phi (\lambda_{j}^{(l)} - \lambda_{k}^{(l)},\gamma)
-\sum_{\alpha=m -2,\pm} \sum_{k=1}^{n_{\alpha}} \Phi (\lambda_{j}^{(l)} - \lambda_{k}^{(\alpha)},\frac{\gamma}{2}) \nonumber \\
\;\;\;\;\;\;\;\;\;\;\;\;\;\;\;\;\;\;\;\;\;\;\;\;\;\;\;\;\;\;\;\;\;\;\;\;\;\;\;\;\;\;\;\;l=m-1 
\label{be2}
\end{eqnarray}
\begin{eqnarray}
\delta_{m,1} L \Phi(\lambda_{j}^{(l)}, \frac{\gamma}{2})=2 \pi Q_{j}^{(\pm)}
+\sum_{k=1}^{n_{\mp}} \Phi (\lambda_{j}^{(l)} - \lambda_{k}^{(\mp)},\gamma) 
-\sum_{k=1}^{n_{m-1}} \Phi (\lambda_{j}^{(l)} - \lambda_{k}^{(m-1)},\frac{ \gamma}{2})  \nonumber \\
\;\;\;\;\;\;\;\;\;\;\;\;\;\;\;\;\;\;\;\;\;\;\;\;\;\;l=\pm
\label{be3}
\end{eqnarray}
where
$\Phi(\lambda,\gamma) = 2 \arctan{\left[ \cot{\left(\gamma \right)}\tanh{\left(\lambda \right)} \right]}$. 
The
numbers $Q_j^{(l)}$ define the different branches of the logarithm and in general are integers or
half-integers. For example, 
part of the low-lying
spectrum can be parameterized in terms of an integer sector index $r_l$ and the corresponding
sequence of $Q_j^{(l)}$ numbers are,
\begin{eqnarray}
Q_{j}^{(l)}&=& -\frac{1}{2} \left[ L-r_{l} -1 \right] + j - 1 \;\;\;\;\;\;\;\; j=1,\dots, L-r_{l} \;\; \mbox{and} \;\; l=1,\dots,m-1 \\
Q_{j}^{(\pm)}&=& -\frac{1}{2} \left[ \frac{L}{2}-r_{\pm} -1 \right] + j - 1 \;\;\;\;\;\;\;\;\;\;\;\;\;\;\;\;\;\;\;\;\;\;\;\;\;\;\;\;\;\;\;\;\;\;\;\;\;\;\;\; j=1,\dots ,\frac{L}{2} - r_{\pm}
\end{eqnarray}

For large $L$, the number of roots $n_{l}$ tends toward a continuous distribution in the real axis and 
the following density of roots can be defined
\begin{equation}
\rho^{(l)}(\lambda^{(l)}) = \lim_{L\rightarrow \infty} \frac{1}{L\left(\lambda_{j+1}^{(l)} -\lambda_{j}^{(l)} \right)}.
\end{equation}

In the limit $ L \rightarrow \infty $, the Bethe ansatz equations (\ref{be1}-\ref{be3}) turn into coupled linear
integral equations for the densities
$\rho^{(l)}(\lambda^{(l)})$. These integral equations can be solved by standard Fourier transform method, 
$\rho^{(l)}(\omega)= \frac{1}{2 \pi}
\displaystyle \int_{-\infty}^{+\infty} e^{i\omega \lambda} \rho^{(l)}(\lambda) d\lambda$, and the final results are
\begin{eqnarray}
\label{s4}
\rho^{(l)}(\omega) = \cases{
\frac{1}{2\pi} \frac{\cosh{\left[(m-l)\frac{\gamma \omega}{2} \right]}}{\cosh{\left[\frac{m \gamma \omega}{2} \right]}} \;\;\;\;\;\;\;\;\;\;\;\;\; l=1,\dots, m-1 \cr
\frac{1}{4\pi} \frac{1}{\cosh{\left[\frac{m\gamma \omega}{2} \right]}} \;\;\;\;\;\;\;\;\;\;\;\;\;\;\;\;\;\; l=\pm }
\end{eqnarray}

From Eqs. (\ref{ene},\ref{s4}) we can compute the ground state energy per site $e_{\infty}^{(m)}(\gamma)$ in the
infinite volume limit. By  writing Eq. (\ref{ene}), in terms of its Fourier transform, we find the expression
\begin{equation}
e_{\infty}^{(m)} (\gamma) = \displaystyle -2 J\int_{0}^{\infty} 
\frac{\sinh{\left[(\pi - \gamma) \frac{\omega}{2} \right]}  
\cosh{\left[ (m-1)\frac{\gamma \omega}{2} \right]}}
{\sinh{\left[ \frac{\pi \omega}{2} \right]} 
\cosh{\left[ m\frac{\gamma \omega}{2} \right]}} d \omega \;\;\;\;\;\;\;\;\;   
\end{equation}

Let us now turn our attention to the behaviour of the low-lying excitations in the thermodynamic limit. As usual
to many integrable models, the energy $\varepsilon^{(l)}(x)$ and the momenta 
$p^{(l)}(x)$  of the $l$-th excitation measured from the ground state are related by 
\begin{equation}
\varepsilon^{(l)}(x) = 2 \pi \rho^{(l)}(x) \;\;\;\;\;\;\;\;\;\; p^{(l)}(x)= \int_{x}^{\infty} 
\varepsilon^{(l)}(y) dy  
\end{equation}

By using Eq. (\ref{s4}) we conclude that the low-momenta dispersion relation is linear for
all the excitations, 
$\varepsilon^{(l)}(p) = v_m(\gamma) p^{(l)}$. The respective sound velocity $v_m(\gamma)$ is found to be
\begin{equation}
v_m(\gamma)= \frac{J \pi}{m \gamma}.
\end{equation}

We have now the basic ingredients to investigate the finite size effects in the spectrum
of the $U_q[(2|2m)]$ spin chain for $0 \leq \gamma \leq \frac{\pi}{2} $.

\section{Finite size properties}

The basic behaviour of the leading finite size corrections to the spectrum
of gapless systems are expected to follow that of conformally invariant theories 
in a strip of width $L$ \cite{CA}.
For periodic boundary conditions, the ground state 
energy $E_0(L)$ behaves, for large L, as
\begin{equation}
\label{e00}
\frac{E_0(L)}{L} =e_{\infty} -\frac{\pi v c}{ 6 L^2} +O(L^{-2}),
\end{equation}
where $c$ is the central charge and $v$ is the sound velocity.

The structure of the higher energy states $E_{\alpha}(L)$ are also determined by the conformal
dimensions $X_{\alpha}$ of the respective primary operators, namely
\begin{equation}
\frac{E_{\alpha}(L)}{L}   
-\frac{E_{0}(L)}{L}   
=\frac{2 \pi v X_{\alpha}}{ L^2} +O(L^{-2}).
\end{equation}

In what follows we begin our study of the finite size effects by considering first
the simplest case $m=1$.

\subsection{The $U_{q}[osp(2|2)]$ model}

The finite size corrections for the $U_q[osp(2|2)]$ spin chain can be studied
with rather little effort at the particular point $\gamma=\frac{\pi}{2}$. In this
case the Bethe ansatz equations for the roots $ \{ \lambda_j^{(\pm)} \}$ become
similar to that of lattice free-fermion models,
\begin{equation}
\left [ \frac{\sinh{\left( \lambda_{j}^{(\pm)} + i\frac{\pi}{4} \right)}}
{\sinh{\left(\lambda_{j}^{(\pm)} - i\frac{\pi}{4} \right)}} \right ]^{L}=
e^{iL k_{j}^{(\pm)}} = (-1)^{n_{\mp}} 
\end{equation}
while the spectrum are parameterized by
\begin{equation}
E^{(1)}( L, \frac{\pi}{2}) = -2J \sum_{j=1}^{n_{+}} \cos{\left( k_{j}^{(+)} \right)} -
2J \sum_{j=1}^{n_{-}} \cos{\left( k_{j}^{(-)} \right)}.
\end{equation}

Therefore, for the value $\gamma=\frac{\pi}{2}$, one can exhibit exact expressions for
the low-lying energies, in a given sector $r_{\pm}$, by summing 
over selected free-momenta of type $k_j^{(\pm)}= \frac{\pi}{L}  \bar{n}_j^{\pm}$ where
$\bar{n}_j^{\pm}$ are integers. The computation depends, however, whether the fermionic
index  $r_{+} +r_{-}$ is an odd or an even number. When $r_{+}+r_{-}$ is an 
odd number we find that
\begin{equation}
E^{(1)}(L,\frac{\pi}{2}) = - 2 J
\frac{\left[ \cos{\left(\frac{\pi r_{+}}{L} \right)} + \cos{\left(\frac{\pi r_{-}}{L} \right)} \right]}{\sin{\left( \frac{\pi}{L} \right)}},
\end{equation}
\n whose asymptotic expansion for large $L$ becomes
\begin{equation}
\label{xx}
\frac{E^{(1)}(L,\frac{\pi}{2})}{L} = e_{\infty}^{(1)} (\frac{\pi}{2}) +
\frac{2\pi}{L^2} v_{1}(\frac{\pi}{2}) \left[ -\frac{1}{6} + \frac{r_{+}^2 + r_{-}^2}{4} \right]
+ O\left( L^{-2} \right).
\end{equation}

Similar analysis can be performed when $r_{+}+r_{-}$ is an even number. The
difference is that now the free-momenta are shifted by a fixed amount $\frac{\pi}{L}$.
For instance, the expression for the lowest energy 
in the sector $r_{+}=r_{-}=0$ is given by
\begin{eqnarray}
\label{xxx}
\frac{E^{(1)}(L,\frac{\pi}{2})}{L} &=& -4J \tan{(\frac{\pi}{L})} \nonumber \\
&=& e_{\infty}^{(1)} (\frac{\pi}{2}) +
\frac{2\pi}{L^2} v_{1}(\frac{\pi}{2}) \left[ -\frac{1}{6} + \frac{1}{2} \right]
+ O\left( L^{-2} \right).
\end{eqnarray}

Direct comparison between Eqs. (\ref{xx},\ref{xxx}) reveals us that the form of the finite size effects
has a clear dependence on the fermionic index $r_{+}+r_{-}$. We also note that the ground state for
finite $L$ lies in the odd sectors $r_{+}=\pm 1$ and $r_{-}=0$ 
or $r_{+}=0$ and $r_{-}=\pm 1$ 
and therefore it is four-fold 
degenerated. This preliminary analysis will be of utility to help us to make a prediction
for the finite size corrections in the whole regime $0 \leq \gamma \leq \frac{\pi}{2}$.

To make further progress  for arbitrary values of the coupling $\gamma$ we use
the so-called  density root method \cite{RO,RO1}. This approach is able to give us
the main expected behaviour of the leading finite size corrections when both the ground
state and the low-lying excitations are described in terms of real roots
or those
carrying a fixed imaginary part such as $i\frac{\pi}{2}$. This is exactly
the situation we found for the $U_q[osp(2|2)]$ model. This conclusion is achieved by comparing
the spectrum
generated by numerical solutions of the Bethe ansatz equations (\ref{fbbf}) 
with that from direct diagonalization of
the Hamiltonian (\ref{ham}) up to $ L =16$. By applying the root density approach 
to the $U_q[osp(2|2)]$  model
one finds that its prediction for the finite size behaviour of the eigenenergies is
\begin{equation}
\label{e22}
\frac{E^{(1)}(L,\gamma)}{L}=  e_{\infty}^{(1)} (\gamma) +
\frac{2\pi}{L^2} v_{1}(\gamma) \left[ -\frac{1}{6} + X_{r_{+},r_{-}}^{s_{+},s_{-}}(\gamma) \right]
+ O\left( L^{-2} \right),
\end{equation}
where the scaling dimensions $\displaystyle X_{r_{+},r_{-}}^{s_{+},s_{-}}(\gamma)$ can be written as
\begin{equation}
\label{quinze}
X_{r_{+},r_{-}}^{s_{+},s_{-}}(\gamma) = (1-\frac{\gamma}{\pi}) \frac{(r_{+} + r_{-})^2}{4} +
\frac{\gamma}{\pi} \frac{(r_{+} - r_{-})^2}{4} 
+\frac{1}{(1-\frac{\gamma}{\pi})} \frac{(s_{+} + s_{-})^2}{4} +
\frac{\pi}{\gamma} \frac{(s_{+} - s_{-})^2}{4} .
\end{equation}

The integers $r_{\pm}$ are related to the number of roots 
$m_{\pm}=\frac{L}{2} - r_{\pm}$
while the numbers $s_{\pm}$ are
directly related to the presence of holes in the $Q_{j}^{(\pm)}$
distribution. The latter indices are rather sensitive to boundary
conditions and therefore they need extra care. In fact, by comparing Eqs. (\ref{e22},\ref{quinze}) at the point
$\gamma=\frac{\pi}{2}$ with Eqs. (\ref{xx},\ref{xxx}) one sees that for $r_{+}+r_{-}$ odd
the number $s_{\pm}$ are expected to start from zero. By way of contrast for $r_{+}=r_{-}=0$ the lowest values
for $s_{\pm}$ is in fact half-integer $\frac{1}{2}$. From our numerical analysis we also conclude
that the standard root density assumptions concerning the values for $s_{\pm}$ are valid only for
anti-periodic boundary conditions in the case $r_{+}+r_{-}$ is an even number. This means that in such
sectors, integers values for $s_{\pm}$ are expected only when a twist $\mathrm{e}^{\pm i \pi}$ multiplies the Bethe
ansatz equations (\ref{fbbf}) for both $\{ \lambda_j^{(\pm)} \}$ variables. It turns out that the effect of a twist
$\mathrm{e}^{i \varphi}$ in the root density method is to shift the numbers 
$s_{\pm}$ by a factor $\frac{\varphi}{2\pi}$. Considering these observations
one concludes that, for periodic boundary conditions, the numbers $s_{\pm}$ should indeed
begin at values $\pm \frac{1}{2}$ when $r_{+}+r_{-}$ is even. These 
arguments strongly suggest that the possible values of 
the vortex numbers $s_{\pm}$ should depend on the
spin-wave numbers $r_{\pm}$ by the following rule,
\begin{eqnarray}
\label{V}
\bullet \;\;\; \mathrm{for} \;\; r_{+}+r_{-} \;\; \mathrm{odd}  \;\;\; &&\rightarrow \;\;\;\;\;  s_{\pm} = 0, \pm 1 ,\pm 2, \dots \nonumber \\
\bullet \;\;\; \mathrm{for} \;\; r_{+}+r_{-} \;\; \mathrm{even} \;\; &&\rightarrow \;\;\;\;\; 
s_{\pm} = \pm \frac{1}{2}, \pm \frac{3}{2} ,\pm \frac{5}{2} , \dots 
\end{eqnarray}

In order to investigate the validity of the proposal (\ref{quinze},\ref{V}) beyond
the decoupling point, we have solved numerically the original Bethe ansatz equations (\ref{fbbf})
for $L \sim 24$. This numerical work enables us to compute the sequence,
\begin{equation}
\label{dezesseis}
X(L) = \left( \frac{E^{(1)}(L)}{L} - e_{\infty}^{(1)}(\gamma) \right) \frac{L^2}{2\pi v_{1}(\gamma)} +
\frac{1}{6},
\end{equation}
\n that are the expected to extrapolate to the dimensions
$\displaystyle X_{r_{+},r_{-}}^{s_{+},s_{-}}(\gamma)$.

In table 1 we show the finite size sequences (\ref{dezesseis}) for some of the
lowest dimensions with $r_{+}+r_{-}=1$ for $\gamma=\frac{\pi}{5},\frac{\pi}{4}$.
The data for $X_{0,1}^{1,1}(\gamma)$ is restricted to $L=16$ due to
numerical instabilities with the respective Bethe roots.
\begin{table}[h]
\begin{center}
\begin{tabular}{|c|c|c|c|c|c|c|} \hline
$L$  &$X_{0,1}^{0,0}(\frac{\pi}{5})$  &$X_{0,1}^{0,1}(\frac{\pi}{5})$ &
$X_{0,1}^{1,1}(\frac{\pi}{5})$ &$X_{0,1}^{0,0}(\frac{\pi}{4})$  &$X_{0,1}^{0,1}(\frac{\pi}{4})$ &
$X_{0,1}^{1,1}(\frac{\pi}{4})$  \\ \hline\hline

8 &0.272854  &1.449160  &1.494484$(*)$ &0.267351  &1.393699  &1.548692$(*)$           \\ \hline
10 &0.269116  &1.504508  &1.510610$(*)$ &0.263877  &1.434436  &1.568194$(*)$          \\ \hline
12 &0.266716  &1.542057  &1.518366$(*)$ &0.261697  &1.460574  &1.578323$(*)$          \\ \hline
14 &0.265022  &1.569286  &1.522338$(*)$ &0.260192  &1.478663  &1.584066$(*)$          \\ \hline
16 &0.263751  &1.590025  &1.524411$(*)$ &0.259085  &1.491841  &1.588751$(*)$          \\ \hline
18 &0.262754  &1.606393  & ----- &0.258233  &1.501900  & -----                      \\ \hline
20 &0.261945  &1.619746 & ----- &0.257555  &1.509764 & -----                        \\ \hline
22 &0.261272  &1.630840 & ----- &0.256999 &1.5161128 & -----                        \\ \hline
24 &0.260702  &1.640206  & -----  &0.256535  &1.521357  & -----                     \\ \hline
Extrap. & 0.250(1)  & 1.811(2) & 1.52(1) & 0.250 0(1)  & 1.581(1) & 1.59(1)
\\ \hline
Exact & 0.25  & 1.812 5 & 1.5 & 0.25  & 1.583 3\dots & 1.583 3\dots
\\ \hline
\end{tabular}
\end{center}
\caption{\footnotesize{Finite size sequences  for the extrapolation
of anomalous dimensions
of the $U_q[Osp(2|2)]$ model for
$\gamma=\pi/5,\pi/4$.
The exact
expected conformal dimensions are $X_{1,0}^{0,0}(\gamma)=1/4$,
$X_{0,1}^{0,1}(\gamma)=1/4+1/[4(\gamma/\pi)(1-\gamma/\pi)]$ and
$X_{0,1}^{1,1}(\gamma)=1/4 +1/(1-\gamma/\pi)$.
The symbol $(*)$ refers to Lanczos numerical data.}}
\end{table}

In Figures $4(a,b)$
we exhibit the pattern of the roots associated to the $X_{0,1}^{0,0}(\gamma)$
and $X_{0,1}^{0,1}(\gamma)$ respectively.

\pagebreak

In table 2 we show similar results for dimensions where $r_{+}+r_{-}$ is
even and the corresponding roots structure are exhibited in
Figures $5(a,b,c)$.
\begin{table}[!h]
\begin{center}
\begin{tabular}{|c|c|c|c|c|c|c|} \hline
$L$  &$X_{0,0}^{\frac{1}{2},\frac{1}{2}}(\frac{\pi}{5})$  &$X_{1,1}^{\frac{1}{2},\frac{1}{2}}(\frac{\pi}{5})$ &
$X_{1,1}^{\frac{1}{2},-\frac{1}{2}}(\frac{\pi}{5})$ &$X_{0,0}^{\frac{1}{2},\frac{1}{2}}(\frac{\pi}{4})$  &$X_{1,1}^{\frac{1}{2},\frac{1}{2}}(\frac{\pi}{4})$ &
$X_{1,1}^{\frac{1}{2},-\frac{1}{2}}(\frac{\pi}{4})$                                     \\ \hline\hline
8 &0.320662  &1.083548  &1.459096 &0.339496  & 1.060728  &1.394641                                              \\ \hline
10 &0.318225  &1.092890  &1.530066 &0.337373 & 1.068928  &1.452543                                             \\ \hline
12 &0.316831  &1.098146  &1.580667 &0.336207 & 1.073385  &1.492308                                             \\ \hline
14 &0.315946  &1.101419  &1.618909 &0.335494 & 1.076069  &1.521388                                               \\ \hline
16 &0.315341  &1.103607  &1.649054 &0.335025 & 1.077806  &1.543652                                             \\ \hline
18 &0.314906  &1.105151  &1.673578 &0.334699 & 1.078993  &1.561296                                             \\ \hline
20 &0.314580  &1.106286 & 1.694025 &0.334462 & 1.079839  &1.575661                                             \\ \hline
22 &0.314327  &1.107148 & 1.711407 &0.334284 & 1.080463  &1.587612                                             \\ \hline
24 &0.314126  &1.107821  &1.726419 &0.334147 & 1.080936  &1.597730                                             \\ \hline
Extrap. & 0.312 53(1)  & 1.112 3(2) & 2.054(1)  & 0.333 34(2)  & 1.833 2(2) & 1.752(1)
\\ \hline
Exact & 0.312 5  & 1.112 5 &  2.05  & 0.333 3\dots  &1.083 3\dots  & 1.75
\\ \hline
\end{tabular}
\end{center}
\caption{\footnotesize{Finite size sequences for the extrapolation
of the  anomalous dimensions
of the $U_q[Osp(2|2)]$ model for
$\gamma=\pi/5,\pi/4$.
The exact  expected conformal dimensions are
$X_{0,0}^{\frac{1}{2},\frac{1}{2}}(\gamma)=1/[4(1-\gamma/\pi)]$,
$X_{1,1}^{\frac{1}{2},\frac{1}{2}}(\gamma)=(1-\gamma/\pi)+1/[4(1-\gamma/\pi)]$ and
$X_{1,1}^{\frac{1}{2},-\frac{1}{2}}(\gamma)=(1-\gamma/\pi)+1/(4\gamma/\pi)$.}}
\end{table}

All these numerical results confirm
the  conjecture (\ref{quinze},\ref{V}) for the finite size properties of
the $U_{q}[osp(2|2)]$ quantum spin chain.
\suppressfloats
We now proceed with a discussion of our results. For periodic boundary conditions the ground
state $E_0^{(1)}(L,\gamma)$ sits in the sectors $r_{+}=\pm 1$ and $r_{-}=0$ or $r_{+}=0$ and $r_{-}=\pm$
and according to the rule (\ref{V}) the respective 
vortex numbers have the lowest possible value $s_{\pm}=0$. 
From Eqs. (\ref{e22},\ref{quinze}) we derive that its
finite size behaviour is,
\begin{equation}
\label{e220}
\frac{E_0^{(1)}(L,\gamma)}{L} = e_{\infty}^{(1)} (\gamma)
+\frac{\pi}{6 L^2} v_{1}(\gamma) +O(L^{-2}).
\end{equation}
\suppressfloats
By comparing Eqs. (\ref{e00},\ref{e220}) we conclude that, in the continuum,
the $U_q[osp(2|2)]$ vertex model should be described in terms of a conformal 
field theory with central charge $c=-1$. The respective dimensions of the
primary operators depends on the anisotropy and measuring them from the
ground state we find that they are  
$X_{r_{+},r_{-}}^{s_{+},s_{-}}(\gamma)-\frac{1}{4}$ where $r_{\pm}$ and $s_{\pm}$
satisfy the condition (\ref{V}). This is probably the first example in the
literature of a theory with  $c < 0$  exhibiting a line of continuously 
varying exponents. In particular, we see the lowest conformal dimension
occurs in the sector $r_{+}=r_{-}=0$ with value  
$X_1= \frac{\gamma}{4 \pi} \frac{1}{(1-\frac{\gamma}{\pi})} $ which degenerates
to that of the ground state for $\gamma=0$.
\suppressfloats
The isotropic point $\gamma=0$ possesses indeed special features. From Eq. (\ref{quinze})
we see that for $s_{+} \neq s_{-}$ the scaling dimensions diverge as $\gamma \rightarrow 0$.
In this limit one expects therefore that only the sectors $s_{+}=s_{-}$ will contribute to the low-energy
operator content. In order to describe the expected scaling dimensions at the isotropic 
point lets us, by considering the rule (\ref{V}), define $s_{+}+s_{-}=2 \bar{s}~(2\bar{s}+1)$ for
$r_{+}+r_{-}=2\bar{r}+1~(2 \bar{r})$.  We see then that the finite part of the dimensions (\ref{quinze})
becomes,
\begin{eqnarray}
\label{iso}
X_{\bar{r}}^{\bar{s}}(0) &=& \cases { \displaystyle \frac{(2 \bar{r}+1)^2}{4} +{\bar{s}}^2 \; ~~~~~r_{+}+r_{-}=2\bar{r}+1 \cr
\displaystyle {\bar{r}}^2 +\frac{(2 \bar{s}+1)^2}{4} ~~~~~~r_{+}+r_{-}=2\bar{r} \cr }
\end{eqnarray}
where $\bar{r},\bar{s}=0,\pm 1, \pm 2, \dots $.

The above conclusions for $\gamma=0$ agree with only part of the recent predictions made in \cite{SAI} for
the possible values of the conformal dimensions of the isotropic $osp(2|2)$ spin chain. 
Although the dimensions (\ref{iso}) are the same for both sectors and coincide with that of a free
boson with radius of compactification $R=1$ or $R=2$ \footnote{ Recall that the conformal dimensions of
a compactified free boson $\phi(x)= \phi(x) +2 \pi R$ are $ \frac{\bar{r}^2 R^2}{4} +\frac{\bar{s}^2} {R^2}$ \cite{GIN}.},
the respective values for the spin-wave or vortex numbers are restricted solely to odd integers.
This subtlety may be of relevance in the description 
of the continuum limit of the $osp(2|2)$ spin chain \cite{SAI1}.

\subsection{The $U_q[osp(2|2m)]$ model}

For arbitrary $ m \geq 2$ the special point $\gamma=\frac{\pi}{2}$ is not of
great help because the Bethe ansatz equations (\ref{fbbf}) are not fully decoupled. 
We shall therefore start our study by considering the analytical predictions
that can be made within the density root method. Considering that such an
approach has already been well described before \cite{RO,RO1}, we will present
here only the final results for general $m$. This framework can be adapted
to handle the nested form of the Bethe equations (\ref{fbbf}) and the finite
size corrections for $E^{(m)}(L,\gamma)$ turns out to be,
\begin{equation}
\label{e2n}
\frac{E^{(m)}(L,\gamma)}{L}=  e_{\infty}^{(m)} (\gamma) +
\frac{2\pi}{L^2} v_{m}(\gamma) \left[ -\frac{(m+1)}{12} + X_{r_1,\dots,r_{m-1},r_{+},r_{-}}^{s_1,\dots,s_{m-1},s_{+},s_{-}}(\gamma) \right]
+ O\left( L^{-2} \right),
\end{equation}
where the corresponding scaling dimensions are given by
\begin{equation}
\label{s5}
X_{r_1,\dots,r_{m-1},r_{+},r_{-}}^{s_1,\dots,s_{m-1},s_{+},s_{-}}(\gamma) =
\frac{1}{4} \sum_{\alpha,\beta=1}^{m-1,\pm} r_{\alpha} C_{\alpha,\beta}^{(m)}(\gamma) r_{\beta} 
+\sum_{\alpha,\beta=1}^{m-1,\pm} s_{\alpha} \left [ C^{(m)}(\gamma) \right ]^{-1}_{\alpha,\beta} s_{\beta},
\end{equation}
and the non-null matrix elements $
C_{\alpha,\beta}^{(m)}(\gamma)$   are
\begin{eqnarray}
\label{s6}
C_{\alpha,\beta}^{(m)}(\gamma) 
&=& \cases{
\displaystyle 2(1-\frac{\gamma}{\pi}) \delta_{\alpha,\beta} -(1-\frac{\gamma}{\pi})\left [ \delta_{\alpha,\beta+1} +\delta_{\alpha,\beta-1}
\right ] \;\;\; \;\;\;\;\;\;\;  \;\; \alpha,\beta=1,\dots,m-1  \cr
\displaystyle -(1-\frac{\gamma}{\pi})  
 \;\;\; \;\;\;\;\;\;\;\;\;\;\;\;\;\;\;\;\; \;\; \alpha=m-1, \;\beta=\pm 
\;\;\; \mathrm{or}\;\; \alpha=\pm, \; \beta=m-1   \cr
\displaystyle  1
  \; \;\;\;\;\;\;\;\;\;\;\;\;\;\;\;\;\; \;\;\;\;\;\;\;\;\;\;\;\;\;\; \;\;\; \alpha=\beta=\pm \;\;\;    \cr
\displaystyle (1-2\frac{\gamma}{\pi}) 
 \;\;\; \;\;\;\;\; \;\;\; \;\;\;\;\;\;\;\;\;\;  \;\; \alpha=\pm, \; \beta=\mp   \cr
 }
\end{eqnarray}

Taking into account previous experience with the $m=1$ case, one would expect the existence of some rule relating the possible values
of the sets
$\{ s_1,\dots,s_{m-1},s_{+},s_{-} \}$  and $ \{ r_1,\dots,r_{m-1},r_{+},r_{-} \}$. 
For $ m \geq 2 $ we encounter some difficulties to unveil possible constraints
between these numbers solely on basis of numerical solutions of the Bethe ansatz equations (\ref{fbbf}) and exact
diagonalization of the respective Hamiltonian (\ref{ham}). In order to make some progress we assume that the origin of 
such rule should go back to the issue of treating strictly periodic boundary conditions for the
fermionic degrees of freedom in all sectors.   We can first consider the situation 
in which all the $2(m+1)$ degrees of freedom behave as bosons as far as boundary conditions are concerned.
Next we look at the sectors whose eigenenergies do change as compared with the original system containing two bosonic
and $2m$ fermionic degrees of freedom.  The sectors whose spectrum remain the same should be described by integers
while the remaining ones by half-integers as far as the values of $s_{\alpha}$ are concerned. Having in mind the above considerations, we are
able to derive the following conjecture for the constraints
\begin{eqnarray}
\label{conj}
\bullet \;\;\;\;\;\; \mathrm{for} \;\; r_{i+1}+r_{i-1} \;\; \mathrm{odd} \;\;\;\;\;\;\;\;&&\rightarrow \;\;\;\;\; 
s_{i} = \pm \frac{1}{2}, \pm \frac{3}{2} ,\pm \frac{5}{2} ,\;  \dots \;\;\;\;\; i=1,\dots,m-2  \nonumber \\
\bullet \;\;\;\;\;\; \mathrm{for} \;\; r_{i+1}+r_{i-1} \;\; \mathrm{even} \;\;\;\;\;\;\;&&\rightarrow \;\;\;\;\; 
s_{i} = 0, \pm 1, \pm 2 ,\; \dots \;\;\;\; \;\;\;\;\;\; i=1,\dots,m-2  \nonumber \\
\bullet \;\;\;\;\;\; \mathrm{for} \;\; r_{m-2}+r_{+} +r_{-}\;\; \mathrm{odd} \;&&\rightarrow \;\;\;\;\; 
s_{m-1} = \pm \frac{1}{2}, \pm \frac{3}{2} ,\pm \frac{5}{2} , \; \dots \nonumber \\
\bullet \;\;\;\;\;\; \mathrm{for} \;\; r_{m-2}+r_{+}+r_{-} \;\; \mathrm{even} &&\rightarrow \;\;\;\;\; 
s_{m-1} = 0, \pm 1, \pm 2 , \; \dots \;\;   \nonumber \\
\bullet \;\;\;\;\;\; \mathrm{for} \;\; r_{m-1}+r_{+}+r_{-} \;\; \mathrm{odd} \;&&\rightarrow \;\;\;\;\; 
s_{\pm} = 0, \pm 1, \pm 2 , \;\dots \;\;  \nonumber \\
\bullet \;\;\;\;\;\; \mathrm{for} \;\; r_{m-1}+r_{+} +r_{-}\;\; \mathrm{even} &&\rightarrow \;\;\;\;\; 
s_{\pm} = \pm \frac{1}{2}, \pm \frac{3}{2} ,\pm \frac{5}{2} , \; \dots    
\end{eqnarray}

Before proceeding we would like to note that the above constraints reflect the structure of the
Bethe ansatz equations (\ref{fbbf}). In fact, the vortex numbers $s_{\alpha}$  depend on the values of the
neighboring spin-wave numbers according to the Dynkin diagram of Figure 3. Furthermore, such relationship
for the Bethe roots with ($\bigcirc$) or without ($\bigotimes$) self-scattering are just the opposite.
In order to give some support to this conjecture we solve numerically the Bethe ansatz
equations (\ref{fbbf}) in the cases $m=2,3$ for some of the low-lying energies. In table 3 we have presented the results
of the extrapolation for three possible dimensions for $m=2$. In table 4 similar data is shown for $m=3$.
\begin{table}[ht]
\begin{center}
\begin{tabular}{|c|c|c|c|c|c|c|} \hline
$L$  &$X_{1,0,0}^{0,0,0}(\frac{\pi}{5})$  &$X_{0,0,0}^{0,\frac{1}{2},\frac{1}{2}}(\frac{\pi}{5})$ &
$X_{1,0,1}^{-\frac{1}{2},\frac{1}{2},\frac{1}{2}}(\frac{\pi}{5})$ &$X_{1,0,0}^{0,0,0}(\frac{\pi}{4})$  &$X_{0,0,0}^{0,\frac{1}{2},\frac{1}{2}}(\frac{\pi}{4})$ &
$X_{1,0,1}^{-\frac{1}{2},\frac{1}{2},\frac{1}{2}}(\frac{\pi}{4})$                               \\ \hline\hline
8 &0.400274  &0.587137  &0.646331  &0.375903  &0.642602  &0.650216                                                      \\ \hline
10 &0.400271  &0.594580  &0.635760 &0.375660  &0.648899  &0.640274                                                      \\ \hline
12 &0.400230  &0.599575  &0.628254 &0.375490  &0.652331  &0.633314                                              \\ \hline
14 &0.400192  &0.603158  &0.622594 &0.375376  &0.655429  &0.628133                                              \\ \hline
16 &0.400160  &0.605863  &0.618139 &0.375296  &0.657300  &0.624103                                             \\ \hline
18 &0.400136  &0.607968  &0.614520 &0.375239  &0.658686  &0.620865                                             \\ \hline
20 &0.400116  &0.609660 &0.6115071 &0.375196  &0.659768 &0.618197                                             \\ \hline
22 &0.400100  &0.611046 &0.608950  &0.375165  &0.660590 &0.615954                                             \\ \hline
24 &0.400088  &0.612220  &0.606745 &0.375140  &0.661270  &0.614038                                             \\ \hline
Extrap. & 0.400 04(1)  & 0.623(3) & 0.562 2(3) & 0.375 1(2)  & 0.664(3) & 0.583 1(3)
\\ \hline
Exact & 0.4  & 0.625 & 0.562 5  & 0.375  &0.666 6\dots  & 0.583 3\dots
\\ \hline
\end{tabular}
\end{center}
\caption{\footnotesize{Finite size sequences  for the extrapolation
of anomalous dimensions
of the $U_q[Osp(2|4)]$ model for
$\gamma=\pi/5,\pi/4$.
The exact
expected conformal dimensions are $X_{1,0,0}^{0,0,0}(\gamma)=(1-\gamma/\pi)/2$,
$X_{0,0,0}^{0,\frac{1}{2},\frac{1}{2}}(\gamma)=1/[2(1-\gamma/\pi)]$ and
$X_{1,0,1}^{-\frac{1}{2},\frac{1}{2},\frac{1}{2}}(\gamma)=1/4 +1/[4(1-\gamma/\pi)]$.}}
\end{table}
\pagebreak
\begin{table}[h]
\begin{center}
\begin{tabular}{|c|c|c|c|c|c|c|} \hline
$L$  &$X_{1,2,1,2}^{0,0,0,0}(\frac{\pi}{5})$  &$X_{1,2,1,1}^{0,\frac{1}{2},-\frac{1}{2},-\frac{1}{2}}(\frac{\pi}{5})$ &
$X_{1,2,0,3}^{0,0,0,0}(\frac{\pi}{5})$
&$X_{1,2,1,2}^{0,0,0,0}(\frac{\pi}{4})$
& $X_{1,2,1,1}^{0,\frac{1}{2},-\frac{1}{2},-\frac{1}{2}}(\frac{\pi}{4})$
& $X_{1,2,0,3}^{0,0,0,0}(\frac{\pi}{4})$
\\ \hline\hline
8 &0.637377    &0.682230   &1.069574          &0.609176   & 0.689721  & 1.109450          \\ \hline
10 &0.643170   &0.700616   &1.067823          &0.615581    & 0.697068 & 1.115921             \\ \hline
12 &0.645843   &0.704425   &1.065523          &0.618706    & 0.700729 & 1.119000                \\ \hline
14 &0.647272   &0.706610   &1.063470          &0.620474    & 0.702871 & 1.120713                \\ \hline
16 &0.648119   &0.708095   &1.061765          &0.621580    & 0.704113 & 1.121774                \\ \hline
18 &0.648658   &0.708884   &1.060367          &0.622321    & 0.705086 & 1.122481                \\ \hline
20 &0.649018   &0.709568   &1.059214          &0.622843    & 0.705583 & 1.122975                \\ \hline
22 &0.649269   &0.710195   &1.058257          &0.623225    & 0.706178 & 1.123337                 \\ \hline
24 &0.649448   &0.710345   &1.057453          &0.623513    & 0.706353 & 1.123608                 \\ \hline
Extrap.  & 0.650 05(1)  & 0.711 8(3) &  1.052 (1)& 0.624 9(2) & 0.708 0(2) & 1.125 2(1)
\\ \hline
Exact & 0.65  &0.712 5  & 1.05 & 0.625 & 0.708 3\dots & 1.125
\\ \hline
\end{tabular}
\end{center}
\caption{\footnotesize{Finite size sequences  for the extrapolation
of anomalous dimensions
of the $U_q[Osp(2|6)]$ model for
$\gamma=\pi/5,\pi/4$.
The exact
expected conformal dimensions are $X_{1,2,1,2}^{0,0,0,0}(\gamma)=(3-2\gamma/\pi)/4$,
$X_{1,2,1,1}^{0,\frac{1}{2},-\frac{1}{2},-\frac{1}{2}}(\gamma)=
(1-\gamma/\pi)/2 +1/[4(1-\gamma/\pi)]$} and
$X_{1,2,0,3}^{0,0,0,0}=(3+6\gamma/\pi)/4$.}
\end{table}

All of them
are in accordance with that predicted by Eqs. (\ref{s5},\ref{s6}) provided the rule (\ref{conj}) is taking into account.

As before, we do not expect that the ground state $E_0^{(m)}(L,\gamma)$ for general $m$ will lie in the sector
with all null spin-wave numbers due to the constraints (\ref{conj}). For $m=2$, combination between Bethe ansatz
and exact diagonalization results leads us to conclude that the ground state sits indeed in the sectors
$r_1=\pm 1$ and $r_{\pm}=0$ or $r_1=r_{+}=r_{-}=\pm 1$. We have verified, for instance, that the lowest energy
in sectors $\{ 1,0,0 \}$ and $\{ 1,1,1 \}$ are exact the same for finite $L$. Consequently, 
from Eqs. (\ref{e2n},\ref{s5}) we derive that
$E_0^{(2)}(L,\gamma)$ behaves as,
\begin{equation}
\label{e2n0}
\frac{E^{(2)}(L,\gamma)}{L} = e_{\infty}^{(m)} (\gamma)- 
\frac{\pi}{6 L^2} v_{2}(\gamma) \left[ 3 -6(1-\frac{\gamma}{\pi}) \right ].
\end{equation}

We see that the term in Eq. (\ref{e2n0}), usually related to the central charge, now
varies continuously with the anisotropy $\gamma$. This is the typical expected behaviour
for the critical properties of loop models \cite{NEE} derived from vertex models with
appropriate boundary conditions such as the $q$-state Potts and six-vertex systems \cite{ALC}.
The criticality of the loop model depends on its fugacity per every loop which turns out
to be a function of the anisotropy $\gamma$ of the corresponding vertex model, see for examples
\cite{OLE}. In our case, strict periodic boundary conditions for both bosonic and
fermionic degrees of freedom should probably work as the bridge between the loop and the
vertex model formulations. This is at least the situation of the isotropic $osp(2|2m)$
vertex model 
which was shown to provide a realization of an intersecting loop model with fugacity
$Q=2(1-m)$ \cite{MB}.  Lets us admit that 
this analogy could be in some manner be extended for
arbitrary $\gamma$. Considering that the $U_q[osp(2|2m)]$ vertex models
share a common underlying braid-monoid algebra \cite{GA1} it is natural  expect that respective loop fugacity will be 
a function of the weight $\bar{Q}$ of  the monoid operator.
From our previous work \cite{GA1} it follows that such weight is
$\bar{Q}= -2 \sin \left [ \gamma(m-1) \right] \frac{\cos(m\gamma)}{\sin(\gamma)} $.  Therefore,
it is only at the special case $m=1$ that $\bar{Q}$ does not depend on the anisotropy, explaining
why in this case the central charge was indeed independent of $\gamma$. A more precise description
of these loop models such as the relation between vertex and loop Boltzmann weights has eluded us
so far.

We have carried out the above analysis up to $m=3$. This leads us to conjecture 
that for general $m$ the finite size correction for the ground state will be,
\begin{equation}
\label{e2nn0}
\frac{E^{(m)}(L,\gamma)}{L} =  e_{\infty}^{(m)} (\gamma)- 
\frac{\pi}{6 L^2} v_{m}(\gamma) \left[ m+1 -3 (m-2[\frac{m}{2}] \frac{\gamma}{\pi}) \right ].
\end{equation}
where $[\frac{m}{2}]$ denotes the largest integer less than $\frac{m}{2}$.

We note that the result (\ref{e2nn0}) when $\gamma \rightarrow 0$ agrees with the central charge
behaviour predicted in \cite{MB} for the isotropic $osp(n|2m)$ spin chains. In this limit, we also
see from Eqs. (\ref{s5},\ref{s6}) that the scaling dimensions for $s_{+} \neq s_{-}$ diverge as
$\gamma$ approaches zero and as before only the sectors $s_{+}=s_{-}$ contributes to the low-lying
operator content. The generality of this scenario for arbitrary $m$ strongly suggests
that the continuum limit of the $osp(2|2m)$ spin chains should be
described by some peculiar field theory. In fact, we remark that a proposal towards this
direction have recently been put forward in the work \cite{SAI1}.

\section{Conclusions}

In this paper we have studied an integrable vertex model invariant relative
to the $U_q[osp(2|2m)]$ quantum superalgebra. The corresponding transfer matrix eigenvalue
problem has been solved by the algebraic Bethe ansatz for a variety of grading choices.
We thus have complemented previous efforts concerning the exact solution of solvable
vertex models based on superalgebras.

We have explored the results for the transfer matrix eigenvalues and Bethe ansatz
equations to investigate the thermodynamic limit properties as well as the finite size corrections
to the spectrum in the massless regime. We have argued that the root density method
needs a subtle adaptation to predict the correct finite size effects. It was observed that
the constraints between spin-wave and vortex numbers  are reflected in the Dynkin representation
of the Bethe ansatz equations. We believe that this will be the general scenario for integrable
models based on superalgebras. This analysis has been helpful to point out possible classes of
universality governing the criticality of the massless phase. The continuum limit of
the $U_q[osp(2|2)]$ vertex model appears to be described by a $c=-1$ conformal theory
with critical exponents varying with the anisotropy. On the other hand, the gapless
regime of the $U_q[osp(2|2m)]$ models for $m \geq 2$ was found to have a multicritical
behaviour typical of loop models of statistical mechanics. 

We hope that our results will open further possibilities of investigations.
For instance, one could use the Bethe ansatz equations (\ref{fbbf}) to
study 
the free-energy thermodynamics
of the $U_{q}[osp(2|2m)]$ vertex models.  This representation is in fact rather
suitable for the application of the so-called quantum transfer matrix method for
finite temperatures \cite{KUP,KUP1}. This would provide us information on relevant
physical properties of the $U_{q}[osp(2|2m)]$ spin chains such as specific heat and
magnetic susceptibility in the entire temperature range. In particular, this could be
used to check the $S$-matrix of a $osp(2|2)$ field theory proposed to described
certain disordered systems \cite{BAZ}.

\section{Acknoledgements}

W. Galleas thanks J.R. Links for useful discussions on orthosympletic algebras and Fapesp
for financial support. M.J Martins thanks Fapesp and CNPq
for partial financial support.

{}

\newpage

\section*{\bf Appendix A: Two-parameter $U_q[osp(2|2)]$ vertex model}
\setcounter{equation}{0}
\renewcommand{\theequation}{A.\arabic{equation}}

In this appendix we present the algebraic Bethe ansatz solution of a two-parameter
$osp(2|2)$ vertex model. These parameters are directly related to the $q$-deformation and
to the continuous $U(1)$ parameter of the four-dimensional representation. The 
respective $\check{R}$-matrix in the $FBBF$ grading can be written as follows,
\begin{eqnarray}
\label{FS}
\check{R}^{(f)}_{12}(\lambda)&=&
\sum_{\alpha=1}^{4} a_{\alpha} (\lambda) \hat{e}_{\alpha \alpha} \otimes \hat{e}_{\alpha \alpha}
+\sum_{\alpha, \beta=1}^{4} d_{\alpha, \beta}(\lambda) \hat{e}_{5-\alpha, \; \beta} \otimes \hat{e}_{\alpha, \; 5-\beta} \nonumber \\
&+& b_{1}(\lambda) \left( \hat{e}_{12} \otimes \hat{e}_{21} + \hat{e}_{21} \otimes \hat{e}_{12} + \hat{e}_{24} \otimes \hat{e}_{42} + \hat{e}_{42} \otimes \hat{e}_{24} \right) \nonumber \\
&+& b_{2}(\lambda) \left( \hat{e}_{13} \otimes \hat{e}_{31} + \hat{e}_{31} \otimes \hat{e}_{13} + \hat{e}_{34} \otimes \hat{e}_{43} + \hat{e}_{43} \otimes \hat{e}_{34} \right) \nonumber \\
&+& c_{1}(\lambda) \left( \hat{e}_{22} \otimes \hat{e}_{11} + \hat{e}_{44} \otimes \hat{e}_{22} \right)
+ c_{2}(\lambda) \left( \hat{e}_{33} \otimes \hat{e}_{11} + \hat{e}_{44} \otimes \hat{e}_{33} \right) \nonumber \\
&+& \bar{c}_{1}(\lambda) \left( \hat{e}_{11} \otimes \hat{e}_{22} + \hat{e}_{22} \otimes \hat{e}_{44} \right)
+ \bar{c}_{2}(\lambda) \left( \hat{e}_{11} \otimes \hat{e}_{33} + \hat{e}_{33} \otimes \hat{e}_{44} \right)
\end{eqnarray}

The main Boltzmann weights are given by 
\begin{eqnarray}
&&a_{1}(\lambda) = -\frac{1}{q_{1}^2} \left( e^{2 \lambda} q_{1}^2 -1 \right)\left( e^{2 \lambda} q_{2}^2 -1 \right)
\;\;\;\;\;\;\;
a_{2}(\lambda) = \frac{1}{q_{1}^2} \left( e^{2 \lambda} q_{1}^2 -1 \right)\left( e^{2 \lambda} - q_{2}^2 \right) \nonumber \\
&&a_{3}(\lambda) = \frac{1}{q_{1}^2} \left( e^{2 \lambda} q_{2}^2 -1 \right)\left( e^{2 \lambda} - q_{1}^2 \right)
\;\;\;\;\;\;\;\;\;\;\;\;
a_{4}(\lambda) = -\frac{1}{q_{1}^2} \left( e^{2 \lambda} q_{1}^2 -1 \right)\left( e^{2 \lambda} q_{2}^2 -1 \right) \nonumber \\
&&b_{1}(\lambda) = \frac{q_{2}}{q_{1}^2} \left( e^{2 \lambda}  -1 \right)\left( e^{2 \lambda} q_{1}^2 -1 \right)
\;\;\;\;\;\;\;\;\;\;\;\;\;\;
b_{2}(\lambda) = \frac{1}{q_{1}} \left( e^{2 \lambda}  -1 \right)\left( e^{2 \lambda} q_{2}^2 -1 \right) \nonumber \\
&&c_{1}(\lambda) = -\frac{1}{q_{1}^2} \left( e^{2 \lambda} q_{1}^2 -1 \right)\left( q_{2}^2 -1 \right)
\;\;\;\;\;\;\;\;\;\;\;\;\;
c_{2}(\lambda) = -\frac{1}{q_{1}^2} \left( e^{2 \lambda} q_{2}^2 -1 \right)\left( q_{1}^2 -1 \right) \nonumber \\
&&\bar{c}_{1}(\lambda) = e^{2\lambda} c_{1}(\lambda)
\;\;\;\;\;\;\;\;\;\;\;\;\;\;\;\;\;\;\;\;\;\;\;\;\;\;\;\;\;\;\;\;\;\;\;\;\;\;\;
\bar{c}_{2}(\lambda) = e^{2\lambda} c_{2}(\lambda)
\end{eqnarray}
\n while the remaining elements can be written as
\begin{eqnarray}
\label{fsb}
d_{\alpha, \beta}(\lambda)=\cases{
-\frac{q_{2}}{q_{1}} \left( e^{2 \lambda} -1 \right)^2 \;\;\;\;\;\;\;\;\;\;\;\;\;\;\;\;\;\;\;\;\;\;\;\;\;\;\;\;\;\;\;\;\;\;\;\;\;\;\;\;\;\;\;\;\;\;\;\;\;\; \alpha=\beta=1,4 \cr
\frac{1}{q_{1}^{2}} \left( e^{2 \lambda} -1 \right) \left( e^{2\lambda} q_{1}^2 q_{2}^2 -1 \right) \;\;\;\;\;\;\;\;\;\;\;\;\;\;\;\;\;\;\;\;\;\;\;\;\;\;\;\;\;\;\;\;\; \alpha=\beta=2,3 \cr
- \frac{1}{q_{1}} \left( e^{2 \lambda} -1 \right) \left( 1 - q_{1}^{-2} \right)^{\frac{1}{2}} \left( 1 - q_{2}^{2} \right)^{\frac{1}{2}}
\;\;\;\;\;\;\;\;\;\;\;\;\;\;\;\;\;\;\;\; \alpha < \beta ,\;\; \beta-\alpha=1,2    \cr
q_{2} e^{2 \lambda} \left( e^{2 \lambda} -1 \right) \left( 1 - q_{1}^{-2} \right)^{\frac{1}{2}} \left( 1 - q_{2}^{2} \right)^{\frac{1}{2}}
\;\;\;\;\;\;\;\;\;\;\;\;\;\;\;\;\;\; \alpha > \beta ,\;\;\; \alpha-\beta=1,2    \cr
-\frac{1}{q_{1}^2} e^{2 \lambda} \left( q_{1}^2 -1 \right) \left( q_{2}^2 -1 \right) \;\;\;\;\;\;\;\;\;\;\;\;\;\;\;\;\;\;\;\;\;\;\;\;\;\;\;\;\;\;\;\;\;\;\;\;\; \alpha=5-\beta=2,3 \cr
\frac{1}{q_{1}^2} \left[ e^{2 \lambda}\left( 1 - q_{1}^2 q_{2}^2 \right) + q_{1}^2 + q_{2}^2 - 2  \right]
\;\;\;\;\;\;\;\;\;\;\;\;\;\;\;\;\;\;\;\;\;\;\;\;\; \alpha=5-\beta=1   \cr
\frac{e^{2 \lambda}}{q_{1}^2} \left[ e^{2 \lambda}\left( q_{1}^2 + q_{2}^2 - 2 q_{1}^2 q_{2}^2 \right) + q_{1}^2 q_{2}^2 - 1 \right]
\;\;\;\;\;\;\;\;\;\;\;\;\;\;\;\;\;\; \alpha=5-\beta=4   \cr}
\end{eqnarray}

It is not difficult to see that 
for $q_{1}=q_{2}=q$ one recovers the $U_{q}[osp(2|2)]$ $R$-matrix defined in Eqs. (\ref{Ryb}-\ref{bwf})
when the grading $FBBF$ is adopted. We start by recalling the definition of  the 
monodromy operator entering the algebraic Bethe ansatz solution of
this vertex model  in the presence of inhomogeneities, 
\begin{equation}
\label{mon}
{\cal T}^{(f)}(\lambda,\{\mu_{j}\})= R^{(f)}_{{\mathcal A} L} (\lambda-\mu_{L})  R^{(f)}_{{\mathcal A} L-1} (\lambda-\mu_{L-1})
\dots R^{(f)}_{{\mathcal A} 1} (\lambda-\mu_{1}),
\end{equation}
\n as well as the associate row-to-row transfer matrix
\begin{equation}
\label{itrans}
T^{(f)}(\lambda,\{\mu_{j}\}) = \mbox{Str} \left[ {\cal T}^{(f)}(\lambda,\{\mu_{j}\}) \right].
\end{equation}

The monodromy operator (\ref{mon}) plays an important role in the formulation of the algebraic
Bethe ansatz method, and with the help of the Yang-Baxter equation one can show that it satisfies the
following quadratic algebra
\begin{equation}
\label{rtt}
\check{R}^{(f)}_{12}(\lambda-\mu)
{\cal T}^{(f)}(\lambda,\{\mu_{j}\})
\stackrel{s}{\otimes}
{\cal T}^{(f)}(\mu,\{\mu_{j}\})=
{\cal T}^{(f)}(\mu,\{\mu_{j}\})
\stackrel{s}{\otimes}
{\cal T}^{(f)}(\lambda,\{\mu_{j}\})
\check{R}^{(f)}_{12}(\lambda-\mu) .
\end{equation}

We remark that the super tensor products in (\ref{rtt}) takes into account the parities in the grading
$FBBF$ \cite{KUL}. Besides that, 
another important ingredient for an algebraic Bethe ansatz solution, 
is the existence of a pseudovacuum state $\ket{\Phi_0}$ in which the monodromy
matrix acts triangularly. For the considered vertex model (\ref{FS}-\ref{fsb}) we can choose
\begin{eqnarray}
\ket{\Phi_0} = \bigotimes_{j=1}^{L} \ket{0}_{j} , ~~
\ket{0}_{j} =
\pmatrix{
1 \cr
0 \cr
0 \cr
0 \cr},
\end{eqnarray}
\n in which the action of the operator $R^{(f)}_{{\cal{A}}j}(\lambda)$ gives
\begin{equation}
\label{acref}
R^{(f)}_{{\cal A}j}(\lambda)\ket{0}_{j} =
\pmatrix{
\omega_1(\lambda) \ket{0}_j &  \dagger  &  \dagger  &   \dagger  \cr
0  &  \omega_2(\lambda) \ket{0}_j &  0  &  \dagger  \cr
0  &  0  &  \omega_{3}(\lambda) \ket{0}_j& \dagger \cr
0  &  0  &  0   &  \omega_{4}(\lambda) \ket{0}_j  \cr}
\end{equation}
The symbol $\dagger$ stands for non-null values while the functions $\omega_{\alpha}(\lambda)$ are given by
\begin{eqnarray}
\omega_{1}(\lambda)&=& - a_{1} (\lambda)  \;\;\;\;\;\;\;\;\;\;\; \omega_{2}(\lambda)= b_{1} (\lambda) \nonumber \\
\omega_{3}(\lambda)&=& b_{2} (\lambda)  \;\;\;\;\;\;\;\;\;\;\;\;\;\; \omega_{4}(\lambda)= -d_{4,4} (\lambda)
\end{eqnarray}

Previous experience with similar vertex models \cite{GA} leads us to adopt the following representation
for the monodromy matrix (\ref{mon})
\begin{equation}
{\cal T}^{(f)}(\lambda,\{\mu_{j}\}) =
\pmatrix{
B(\lambda, \{\mu_{j}\}) &  B_{1}(\lambda, \{\mu_{j}\}) & B_{2}(\lambda, \{\mu_{j}\})  &   F(\lambda ,\{\mu_{j}\})   \cr
C_{1}(\lambda , \{\mu_{j}\})  &  A_{11}(\lambda , \{\mu_{j}\}) &  A_{12}(\lambda , \{\mu_{j}\})   &  B^{*}_{1}(\lambda , \{\mu_{j}\})   \cr
C_{2}(\lambda , \{\mu_{j}\})  &  A_{21}(\lambda , \{\mu_{j}\}) &  A_{22}(\lambda , \{\mu_{j}\})   &  B^{*}_{2}(\lambda , \{\mu_{j}\})   \cr
C(\lambda , \{\mu_{j}\})  & C^{*}_{1}(\lambda , \{\mu_{j}\}) & C^{*}_{2}(\lambda , \{\mu_{j}\}) &  D(\lambda , \{\mu_{j}\})  \cr},
\end{equation}
\n and the diagonalization problem for the transfer matrix becomes equivalent to the problem,
\begin{equation}
\left[ - B(\lambda , \{\mu_{j}\}) +\sum_{i=1}^{2} \hat{A}_{ii}(\lambda, \{\mu_{j}\}) - D(\lambda , \{\mu_{j}\}) \right]
\ket{\phi} =\Lambda^{(f)}(\lambda , \{\mu_{j}\} ) \ket{\phi}.
\end{equation}

The triangular form exhibited by (\ref{acref}) together with (\ref{mon}) allow us to compute
the action of elements of the monodromy matrix ${\cal T}^{(f)}(\lambda,\{\mu_{j}\}) $
on the pseudovaccum state $\ket{\Phi_0}$. In this way we can
regard $B_{1}(\lambda, \{\mu_{j}\})$, $B_{2}(\lambda, \{\mu_{j}\})$ and $F(\lambda ,\{\mu_{j}\})$
as creation fields while the diagonal ones satisfy the relations
\begin{eqnarray}
\label{V1}
B(\lambda , \{\mu_{j}\})\ket{\Phi_0} &=& \prod_{i=1}^{L} \omega_1(\lambda - \mu_{i}) \ket{\Phi_0}  \;\;\;\;\;\;\;\; D(\lambda, \{\mu_{j}\})\ket{\Phi_0} =
\prod_{i=1}^{L} \omega_4(\lambda - \mu_{i}) \ket{\Phi_0} \nonumber \\
A_{11}(\lambda , \{\mu_{j}\} ) \ket{\Phi_0} &=& \prod_{i=1}^{L} \omega_2(\lambda - \mu_{i}) \ket{\Phi_0}  \;\;\;\;\;\;
A_{22}(\lambda , \{\mu_{j}\} ) \ket{\Phi_0} = \prod_{i=1}^{L} \omega_3(\lambda - \mu_{i}) \ket{\Phi_0}, \nonumber \\
\end{eqnarray}
\n as well as annihilation properties for the remaining elements
\begin{eqnarray}
\label{V2}
C(\lambda ,\{\mu_{j}\})\ket{\Phi_0} &=& 0 \;\;\;\;\;\;\;\;\;\;\;
A_{ij}(\lambda ,\{\mu_{j}\} )\ket{\Phi_0} = 0 \;\;\;\;\;\; i \neq j \nonumber \\
C_{i}(\lambda ,\{\mu_{j}\})\ket{\Phi_0} &=& 0  \;\;\;\;\;\;\;\;\;\;\;\;
C_{i}^{*}(\lambda ,\{\mu_{j}\})\ket{\Phi_0} = 0
\end{eqnarray}
The above relations imply that $\ket{\Phi_0}$ is an eigenstate of the transfer matrix whose respective eigenvalue is
\begin{eqnarray}
\label{V3}
\Lambda_{0}^{(f)}(\lambda)=
-\prod_{i=1}^{L} \omega_1(\lambda - \mu_{i})
+\prod_{i=1}^{L} \omega_2(\lambda - \mu_{i})
+\prod_{i=1}^{L} \omega_3(\lambda - \mu_{i})
-\prod_{i=1}^{L} \omega_4(\lambda - \mu_{i}).
\end{eqnarray}

Within the algebraic Bethe ansatz method we now look for the remaining transfer matrix eigenvectors as
linear combinations of products of creation fields acting on $\ket{\Phi_0}$. The general form of these
eigenvectors has been already presented in \cite{GA}. In order to accomplish that we need to
disentangle from Yang-Baxter algebra (\ref{rtt}) appropriate commutation rules between the diagonal and
creation fields. Until this stage, this approach is quite similar to the one used in \cite{GA} unless by
the fact that the presence of two deformation parameters modifies the set of commutation
rules required. In order to avoid an overcrowded section, these commutation rules
have been collected in appendix B.

A careful analysis of the commutation rules given in appendix B, together with the relations
(\ref{V1},\ref{V2},\ref{V3}), leave us with
\begin{eqnarray}
\Lambda^{(f)}( \lambda , \{\mu_{j}\}) &=&
-\prod_{i=1}^{L} \omega_1(\lambda - \mu_{i})
\prod_{i=1}^{n} - a_{1}(\lambda_{i} - \lambda) \; \Lambda_{B}(\lambda, \{ \lambda_{i} \})
+\Lambda_{A}(\lambda, \{ \lambda_{i} \}) \nonumber \\
&& - \prod_{i=1}^{L} \omega_4(\lambda - \mu_{i})
\prod_{i=1}^{n} - \frac{1}{d_{4,4}(\lambda - \lambda_{i})} \; \Lambda_{D}(\lambda, \{ \lambda_{i} \})
\end{eqnarray}

\n where $\Lambda_{B}(\lambda, \{ \lambda_{i} \})$,
$\Lambda_{D}(\lambda, \{ \lambda_{i} \})$ and $\Lambda_{A}(\lambda, \{ \lambda_{i} \})$
are eigenvalues of the auxiliary matrices
$T_{B}(\lambda, \{ \lambda_{i} \})$, $T_{D}(\lambda, \{ \lambda_{i} \})$
and $T_{A}(\lambda, \{ \lambda_{i} \})$
respectively. The
set of rapidities $\{ \lambda_{j} \}$ follows from the vanishing condition of the so called unwanted terms which will be
discussed later.

Initially we shall consider the auxiliary transfer matrix $T_{A}(\lambda, \{ \lambda_{i} \})$ defined as
\begin{eqnarray}
T_{A}(\lambda, \{ \lambda_{i} \})=
\mbox{Tr} \left[ G (\lambda, \{ \lambda_{i} \}) r_{a 1}(\lambda - \lambda_{1})
r_{a 2}(\lambda - \lambda_{2}) \dots r_{a n}(\lambda - \lambda_{n}) \right ]
\end{eqnarray}
\n with the following structure for the auxiliary $r$-matrix
\begin{equation}
\label{ra}
r(\lambda)=\pmatrix{
a_{1}^{*} (\lambda) & 0 & 0 & 0 \cr
0 & b^{*} (\lambda) & 0 & 0  \cr
0 & 0 & b^{*} (\lambda) & 0 \cr
0 & 0 & 0 & a_{2}^{*} (\lambda) \cr}.
\end{equation}
\n The corresponding Boltzmann weights are
\begin{eqnarray}
a_{1}^{*}(\lambda) &=& \frac{1}{q_{1}^2} \left( e^{2 \lambda} q_{1}^2 -1 \right) \left( e^{2 \lambda} - q_{2}^2 \right) \nonumber \\
a_{2}^{*}(\lambda) &=& \frac{1}{q_{1}^2} \left( e^{2 \lambda} q_{2}^2 -1 \right) \left( e^{2 \lambda} - q_{1}^2 \right) \nonumber \\
b^{*}(\lambda) &=& \frac{1}{q_{1}^2} \left( e^{2 \lambda} q_{1}^2 -1 \right) \left( e^{2 \lambda} q_{2}^2 -1 \right)
\end{eqnarray}
\n and the diagonal twist is given by
\begin{equation}
G (\lambda, \{ \lambda_{i} \})=\pmatrix{
\displaystyle \frac{\prod_{i=1}^{L} \omega_{2}(\lambda - \mu_{i})}{\prod_{i=1}^{n} b_{1}(\lambda - \lambda_{i})} & 0 \cr
0 & \displaystyle \frac{\prod_{i=1}^{L} \omega_{3}(\lambda - \mu_{i})}{\prod_{i=1}^{n} b_{2}(\lambda - \lambda_{i})} \cr}.
\end{equation}

We are interested in the solution of the eigenvalue problem
\begin{equation}
T_{A}(\lambda, \{ \lambda_{i} \}) \vec{\mathcal{F}} = \Lambda_{A}(\lambda, \{ \lambda_{i} \}) \vec{\mathcal{F}}
\end{equation}
\n which is trivial due to the diagonal form of (\ref{ra}).

Defining the spin up state $\ket{\uparrow}=\pmatrix{1 \cr 0 \cr}$ and the spin down state $\ket{\downarrow}=\pmatrix{0 \cr 1 \cr}$, we can write
\begin{equation}
\label{VF}
\vec{\mathcal{F}}= \ket{\uparrow} \otimes \ket{\uparrow} \otimes \dots
\ket{\downarrow} \otimes \dots \otimes \ket{\uparrow}
\end{equation}

\n possessing $n_{+}$ spin up states and $n_{-}$ spin down states such that $n=n_{+}+n_{-}$. In this basis
it is convenient to separate the set of rapidities $\{ \lambda_{j} \}$ into two subsets $\{ \lambda_{j}^{+} \}$ and $\{ \lambda_{j}^{-} \}$, each one associated
with the spin up and spin down components of $\vec{\mathcal{F}}$ respectively. With the above considerations we have
\begin{eqnarray}
\Lambda_{A}(\lambda, \{ \lambda_{i} \}) &=&
\prod_{i=1}^{L} \omega_{2}(\lambda - \mu_{i})
\prod_{i=1}^{n_{+}} \frac{a_{1}^{*}(\lambda - \lambda_{i}^{+})}{b_{1}(\lambda - \lambda_{i}^{+})}
\prod_{i=1}^{n_{-}} \frac{b^{*}(\lambda - \lambda_{i}^{-})}{b_{1}(\lambda - \lambda_{i}^{-})} \nonumber \\
&+& \prod_{i=1}^{L} \omega_{3}(\lambda - \mu_{i})
\prod_{i=1}^{n_{+}} \frac{b^{*}(\lambda - \lambda_{i}^{+})}{b_{2}(\lambda - \lambda_{i}^{+})}
\prod_{i=1}^{n_{-}} \frac{a_{2}^{*}(\lambda - \lambda_{i}^{-})}{b_{2}(\lambda - \lambda_{i}^{-})}
\end{eqnarray}

Next we turn to the auxiliary matrices $T_{B}(\lambda, \{ \lambda_{i} \})$
and $T_{D}(\lambda, \{ \lambda_{i} \})$, and their respective eigenvalues. By way of contrast,
these matrices are not defined as a trace of a monodromy matrix, but they are diagonal matrices whose elements
are given by
\begin{eqnarray}
T_{B}(\lambda, \{ \lambda_{i} \})^{\alpha_{1} \alpha_{2} \dots \alpha_{n}}_{\beta_{1} \beta_{2} \dots \beta_{n}}&=&
\prod_{i=1}^{n} \frac{1}{b_{\alpha_{i}}(\lambda_{i} - \lambda)} \delta_{\alpha_{1} \beta_{1}} \delta_{\alpha_{2} \beta_{2}}
\dots \delta_{\alpha_{n} \beta_{n}} \nonumber \\
T_{D}(\lambda, \{ \lambda_{i} \})^{\alpha_{1} \alpha_{2} \dots \alpha_{n}}_{\beta_{1} \beta_{2} \dots \beta_{n}}&=&
\prod_{i=1}^{n} b_{\alpha_{i}}(\lambda - \lambda_{i}) \delta_{\alpha_{1} \beta_{1}} \delta_{\alpha_{2} \beta_{2}}
\dots \delta_{\alpha_{n} \beta_{n}}
\end{eqnarray}

Considering the trivial eigenvectors $\vec{\mathcal{F}}$ (\ref{VF}) we are left with
\begin{eqnarray}
\Lambda_{B}(\lambda, \{ \lambda_{i} \}) &=& \prod_{i=1}^{n_{+}} \frac{1}{b_{1}(\lambda_{i}^{+} - \lambda)}
\prod_{i=1}^{n_{-}} \frac{1}{b_{2}(\lambda_{i}^{-} - \lambda)} \nonumber \\
\Lambda_{D}(\lambda, \{ \lambda_{i} \}) &=& \prod_{i=1}^{n_{+}} b_{1}(\lambda - \lambda_{i}^{+})
\prod_{i=1}^{n_{-}} b_{2}(\lambda - \lambda_{i}^{-})
\end{eqnarray}

In this algebraic Bethe ansatz construction the unwanted terms are canceled out
by making use of explicit form for $\vec{\mathcal{F}}$ (\ref{VF}) and provided that the set of rapidities
$\{ \lambda_{i}^{+} \}$ and $\{ \lambda_{i}^{-} \}$ satisfy suitable Bethe ansatz equations.
Putting our results together, the eigenvalues $\Lambda^{(f)}(\lambda, \{ \mu_{i} \})$ are
\begin{eqnarray}
\label{VAI1}
\Lambda^{(f)}(\lambda, \{ \mu_{i} \})=
&-& \prod_{i=1}^{L} \omega_{1}(\lambda - \mu_{i})
\frac{Q_{+}\left( \lambda - i\frac{\gamma_{2}}{2}\right)}{Q_{+}\left( \lambda + i\frac{\gamma_{2}}{2}\right)}
\frac{Q_{-}\left( \lambda - i\frac{\gamma_{1}}{2}\right)}{Q_{-}\left( \lambda + i\frac{\gamma_{1}}{2}\right)} \nonumber \\
&+& \prod_{i=1}^{L} \omega_{2}(\lambda - \mu_{i})
\frac{Q_{+}\left( \lambda - i\frac{\gamma_{2}}{2}\right)}{Q_{+}\left( \lambda + i\frac{\gamma_{2}}{2}\right)}
\frac{Q_{-}\left( \lambda + i\gamma_{2} + i\frac{\gamma_{1}}{2} \right)}{Q_{-}\left( \lambda + i\frac{\gamma_{1}}{2} \right)} \nonumber \\
&+& \prod_{i=1}^{L} \omega_{3}(\lambda - \mu_{i})
\frac{Q_{+}\left( \lambda + i\gamma_{1} + i \frac{\gamma_{2}}{2} \right)}{Q_{+}\left( \lambda + i\frac{\gamma_{2}}{2} \right)}
\frac{Q_{-}\left( \lambda - i\frac{\gamma_{1}}{2}\right)}{Q_{-}\left( \lambda + i\frac{\gamma_{1}}{2}\right)}  \nonumber \\
&-& \prod_{i=1}^{L} \omega_{4}(\lambda - \mu_{i})
\frac{Q_{+}\left( \lambda + i\gamma_{1} + i \frac{\gamma_{2}}{2} \right)}{Q_{+}\left( \lambda + i\frac{\gamma_{2}}{2} \right)}
\frac{Q_{-}\left( \lambda + i\gamma_{2} + i\frac{\gamma_{1}}{2} \right)}{Q_{-}\left( \lambda + i\frac{\gamma_{1}}{2} \right)}, \nonumber \\
\end{eqnarray}
where as in the main text $Q_{\pm}(\lambda) =\displaystyle \prod_{i=1}^{n_{\pm}} \sinh{\left( \lambda-\lambda_i^{(\pm)} \right)}$.

The corresponding Bethe ansatz equations for the variable $\{ \lambda_j^{(\pm)} \}$ are given by,
\begin{eqnarray}
\label{VAI2}
\prod_{i=1}^{L} \frac{\sinh{\left( \lambda_{j}^{+} - \mu_{i} + i\frac{\gamma_{2}}{2} \right)}}{\sinh{\left( \lambda_{j}^{+} - \mu_{i} - i\frac{\gamma_{2}}{2} \right)}}&=&
\prod_{i=1}^{n_{-}} \frac{\sinh{\left( \lambda_{j}^{+} - \lambda_{i}^{-} + i \frac{(\gamma_{1} + \gamma_{2})}{2} \right)}}{\sinh{\left( \lambda_{j}^{+} - \lambda_{i}^{-} - i\frac{(\gamma_{1}+\gamma_{2})}{2} \right)}} \nonumber \\
\prod_{i=1}^{L} \frac{\sinh{\left( \lambda_{j}^{-} - \mu_{i} + i\frac{\gamma_{1}}{2} \right)}}{\sinh{\left( \lambda_{j}^{-} - \mu_{i} - i\frac{\gamma_{1}}{2} \right)}}&=&
\prod_{i=1}^{n_{+}} \frac{\sinh{\left( \lambda_{j}^{-} - \lambda_{i}^{+} + i \frac{(\gamma_{1} + \gamma_{2})}{2} \right)}}{\sinh{\left( \lambda_{j}^{-} - \lambda_{i}^{+} - i\frac{(\gamma_{1}+\gamma_{2})}{2} \right)}} .
\end{eqnarray}

We finally remark that in the above relations we have set $q_{1,2}=e^{i\gamma_{1,2}}$ and considered the shifts
$\{ \lambda_{j}^{+} \} \rightarrow \{ \lambda_{j}^{+} \} - i\frac{\gamma_{2}}{2}$ and
$\{ \lambda_{j}^{-} \} \rightarrow \{ \lambda_{j}^{-} \} - i\frac{\gamma_{1}}{2}$. In order to obtain the nested
Bethe ansatz solution of the $U_q[osp(2|2m)]$ vertex models one has to use Eqs.(\ref{VAI1},\ref{VAI2}) in the
final step of the recurrence relations (\ref{eirec},\ref{beterec}) at the particular point $\gamma_1=\gamma_2=\gamma$.

\section*{\bf Appendix B}
\setcounter{equation}{0}
\renewcommand{\theequation}{B.\arabic{equation}}

In this appendix we have collected the set of commutation rules required to perform the algebraic
Bethe ansatz for the two parameters $U_{q}[osp(2|2)]$ presented in appendix A.

\begin{eqnarray}
B(\lambda)  B_{i}(\mu)&=& -\frac{a_{1}(\mu-\lambda)}{b_{i}(\mu-\lambda)} B_{i}(\mu)  B(\lambda)
+ \frac{c_{i}(\mu-\lambda)}{b_{i}(\mu-\lambda)}  B_{i} (\lambda) B(\mu)  \\
D(\lambda)  B_{i}(\mu)&=& -\frac{b_{i}(\lambda-\mu)}{d_{4,4}(\lambda-\mu)}  B_{i}(\mu)  D(\lambda)
- \frac{d_{4,1}(\lambda-\mu)}{d_{4,4}(\lambda-\mu)} F(\lambda) C_{i}^{*}(\mu) \nonumber \\
&+& \frac{c_{i}(\lambda-\mu)}{d_{4,4}(\lambda-\mu)} F(\mu) C_{i}^{*}(\lambda)
- \frac{d_{4,j+1}(\lambda-\mu)}{d_{4,4}(\lambda-\mu)} \delta_{j,3-k}
B_{j}^{*}(\lambda) A_{k i}(\mu)
\end{eqnarray}
\begin{eqnarray}
A_{ij}(\lambda) B_{k}(\mu) &=& \frac{1}{b_{i}(\lambda-\mu)} B_{l}(\mu) A_{im}(\lambda) r_{lm}^{jk}(\lambda-\mu)
- \frac{\bar{c_{i}}(\lambda-\mu)}{b_{i}(\lambda-\mu)}  B_{j}(\lambda) A_{ik}(\mu) \nonumber \\
&-&\frac{d_{4,j+1}(\lambda - \mu)}{d_{4,4}(\lambda-\mu)} \delta_{j, 3-k}  B_{i}^{*}(\lambda) B(\mu) +  \frac{d_{4,j+1}(\lambda - \mu)}{d_{4,4}(\lambda - \mu)} \frac{\bar{c_{i}}(\lambda-\mu)}{b_{i}(\lambda-\mu)}
\delta_{j,3-k}  F(\lambda) C_{i}(\mu) \nonumber \\
&+&\frac{1}{b_{i}(\lambda-\mu)} \left[ d_{1,j+1}(\lambda-\mu)  - \frac{d_{4,j+1}(\lambda-\mu) d_{1,4}(\lambda-\mu)}{d_{4,4}(\lambda-\mu)}  \right]
\delta_{j,3-k} F(\mu)  C_{i}(\lambda) \nonumber \\
\end{eqnarray}

\begin{eqnarray}
B(\mu)  F(\lambda)&=& \frac{a_{1}(\lambda -\mu)}{d_{4,4} (\lambda -\mu)} F(\lambda) B(\mu) -
\frac{d_{1,4} (\lambda -\mu)}{d_{4,4} (\lambda -\mu)} F(\mu)  B(\lambda) \nonumber \\
&+& \frac{d_{j+1,4}(\lambda -\mu)}{d_{4,4} (\lambda -\mu)} \delta_{3-i,j} B_{i}(\mu) B_{j}(\lambda) \\
D(\lambda)  F(\mu) &=& \frac{a_{4}(\lambda -\mu)}{d_{4,4} (\lambda -\mu)} F(\mu)  D(\lambda) - \frac{d_{4,1} (\lambda -\mu)}{d_{4,4} (\lambda -\mu)} F(\lambda)  D(\mu) \nonumber \\
&-&\frac{d_{4,i+1}(\lambda -\mu)}{d_{4,4}(\lambda -\mu)} \delta_{i,3-j} B_{i}^{*}(\lambda) B_{j}^{*}(\mu)
\end{eqnarray}
\begin{eqnarray}
A_{ij}(\lambda)  F(\mu) &=& \left[
\frac{b_{j}(\lambda -\mu)}{b_{i} (\lambda -\mu)} -
\frac{c_{j}(\lambda -\mu)}{b_{i} (\lambda -\mu)} \frac{\bar{c_{j}}(\lambda -\mu)}{b_{j} (\lambda -\mu)} \right]
F(\mu) A_{ij}(\lambda) + \frac{\bar{c_{i}}(\lambda -\mu)}{b_{i} (\lambda -\mu)} \frac{\bar{c_{j}}(\lambda -\mu)}{b_{j} (\lambda -\mu)}
F(\lambda) A_{ij}(\mu) \nonumber \\
&-& \frac{\bar{c_{i}}(\lambda -\mu)}{b_{i} (\lambda -\mu)} B_{j}(\lambda) B_{i}^{*}(\mu)
-  \frac{\bar{c_{j}}(\lambda -\mu)}{b_{j} (\lambda -\mu)}  B_{i}^{*}(\lambda) B_{j}(\mu)
\end{eqnarray}
\begin{eqnarray}
B_{i}(\lambda) B_{j}(\mu) &=& \frac{1}{a_{1} (\lambda -\mu)} B_{k}(\mu) B_{l}(\lambda) r_{kl}^{ij} (\lambda -\mu)
-  \frac{d_{4,i+1}(\lambda -\mu)}{d_{4,4} (\lambda -\mu)} \delta_{i,3-j} F(\lambda) B(\mu) \nonumber \\
&-& \frac{1}{a_{1} (\lambda -\mu)} \left[ d_{1,i+1}(\lambda-\mu)  - \frac{d_{4,i+1}(\lambda-\mu) d_{1,4}(\lambda-\mu)}{d_{4,4}(\lambda-\mu)}  \right]
\delta_{i,3-j} F(\mu) B(\lambda) \nonumber \\
\end{eqnarray}
\begin{eqnarray}
\left[ F(\lambda), F(\mu) \right] = 0
\end{eqnarray}
\begin{eqnarray}
F(\mu) B_{i}(\lambda) = - \frac{a_{1} (\lambda -\mu)}{b_{i}(\lambda-\mu)} B_{i}(\lambda) F(\mu)
+ \frac{\bar{c_{i}}(\lambda -\mu)}{b_{i} (\lambda -\mu)} B_{i}(\mu) F(\lambda)
\end{eqnarray}
\begin{eqnarray}
B_{i}(\mu) F(\lambda) = - \frac{a_{1} (\lambda -\mu)}{b_{i}(\lambda-\mu)} F(\lambda) B_{i}(\mu)
+ \frac{c_{i}(\lambda -\mu)}{b_{i} (\lambda -\mu)} F(\mu) B_{i}(\lambda)
\end{eqnarray}
\begin{eqnarray}
B(\mu) B_{i}^{*}(\lambda) &=& - \frac{b_{i}(\lambda- \mu)}{d_{4,4}(\lambda-\mu)} B_{i}^{*}(\lambda) B(\mu)
- \frac{d_{1,4}(\lambda -\mu)}{d_{4,4} (\lambda -\mu)} F(\mu) C_{i}(\lambda) \nonumber \\
&+& \frac{\bar{c_{i}}(\lambda -\mu)}{d_{4,4} (\lambda -\mu)}  F(\lambda) C_{i}(\mu)
- \frac{d_{k+1,4}(\lambda -\mu)}{d_{4,4} (\lambda -\mu)} \delta_{3-i,k} B_{i}(\mu) A_{jk}(\lambda) \\
B_{i}(\mu) B_{j}^{*}(\lambda) &=& - \frac{b_{j}(\lambda -\mu)}{b_{i} (\lambda -\mu)} B_{j}^{*}(\lambda) B_{i}(\mu)
+ \frac{\bar{c_{j}}(\lambda -\mu)}{b_{i} (\lambda -\mu)} F(\lambda) A_{ji}(\mu) \nonumber \\
&-& \frac{c_{i}(\lambda -\mu)}{b_{i} (\lambda -\mu)} F(\mu) A_{ji}(\lambda)
\end{eqnarray}
\begin{eqnarray}
C_{i}(\lambda) B_{j}(\mu) &=& -\frac{b_{j}(\lambda -\mu)}{b_{i} (\lambda -\mu)} B_{j}(\mu) C_{i}(\lambda)
- \frac{\bar{c}_{j}(\lambda -\mu)}{b_{i} (\lambda -\mu)} B(\mu) A_{ij}(\lambda) \nonumber \\
&+& \frac{\bar{c}_{i}(\lambda -\mu)}{b_{i} (\lambda -\mu)} B(\lambda) A_{ij}(\mu) \\
C_{i}^{*}(\lambda) B_{j}(\mu) &=& \frac{1}{d_{4,4}(\lambda-\mu)} B_{k}(\mu) C_{l}^{*}(\lambda)\mathcal{R}_{kl}^{ij}(\lambda-\mu)
-\frac{d_{4,1}(\lambda -\mu)}{d_{4,4} (\lambda -\mu)} B_{j}(\lambda) C_{i}^{*}(\mu) \nonumber \\
&+& \frac{d_{4,k+1}(\lambda -\mu)}{d_{4,4} (\lambda -\mu)} \delta_{k,3-l} A_{ki}(\lambda) A_{lj}(\mu)
- \frac{d_{1,i+1}(\lambda -\mu)}{d_{4,4} (\lambda -\mu)} \delta_{i,3-j} F(\mu) C(\lambda) \nonumber \\
&-& \frac{d_{4,i+1}(\lambda -\mu)}{d_{4,4} (\lambda -\mu)} \delta_{i,3-j} B(\mu) D(\lambda)
\end{eqnarray}

In order to clarify our notation, the elements
$r_{kl}^{ij}$ are obtained from (\ref{ra}) through the definition
\begin{equation}
r(\lambda)=\displaystyle \sum_{i,j,k,l}^{2} r_{ki}^{jl}(\lambda)
\hat{e}_{ij} \otimes \hat{e}_{kl}.
\end{equation}
Finally, we have also used the relation $\mathcal{R}_{k,l}^{i,j}(\lambda)=\check{R}_{k+1,l+1}^{i+1,j+1}(\lambda)$
where the elements $\check{R}_{k,l}^{i,j}$ follows from the convention
\begin{equation}
\check{R}(\lambda) = \sum_{i,j,k,l}^{4} \check{R}_{j,l}^{i,k}(\lambda) \hat{e}_{ij} \otimes \hat{e}_{kl}.
\end{equation}

\newpage 

\begin{figure}[ht]
\begin{center}
\includegraphics{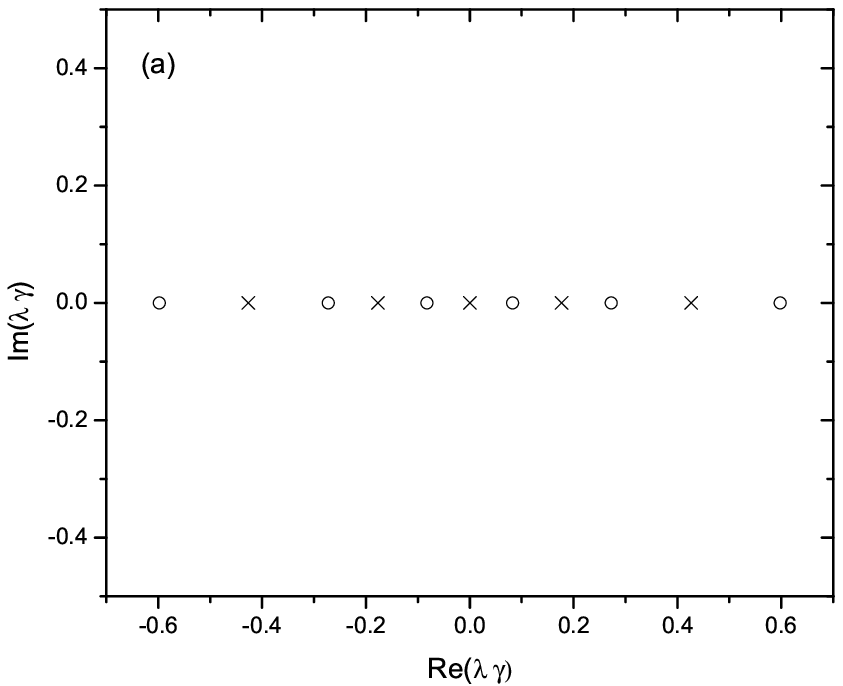}
\end{center}
\begin{center}
\includegraphics{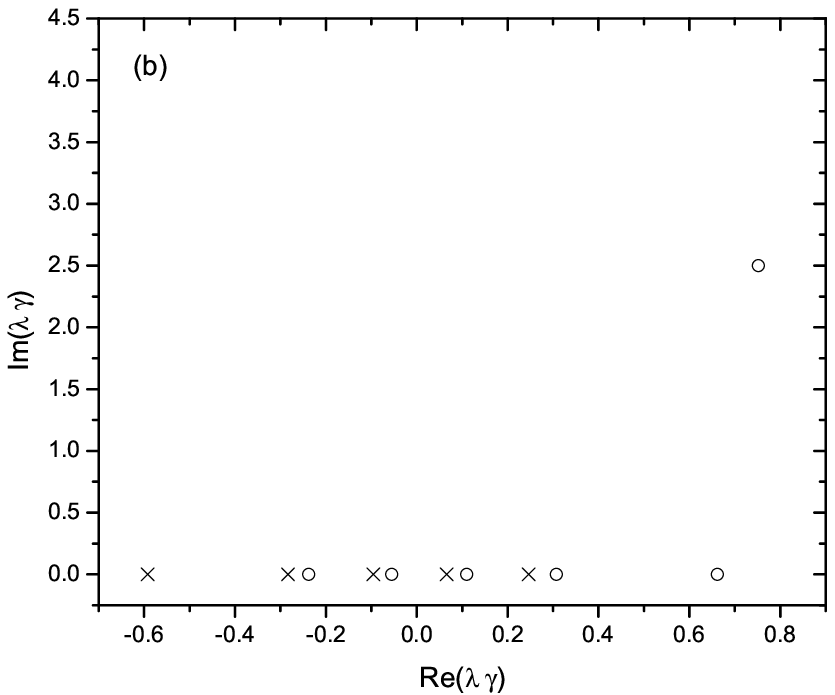}
\end{center}
\caption{\footnotesize{
The Bethe ansatz roots $\lambda_j^{(+)}$ ($\circ$) and
$\lambda_j^{(-)}$ ($\times$) for $\gamma=\frac{\pi}{5}$ and $L=12$. The roots refer
to the dimensions (a) $X_{0,1}^{0,0}(\gamma)$ and (b) $X_{0,1}^{0,1}(\gamma)$.}}
\end{figure}
\begin{figure}[ht]
\begin{center}
\includegraphics{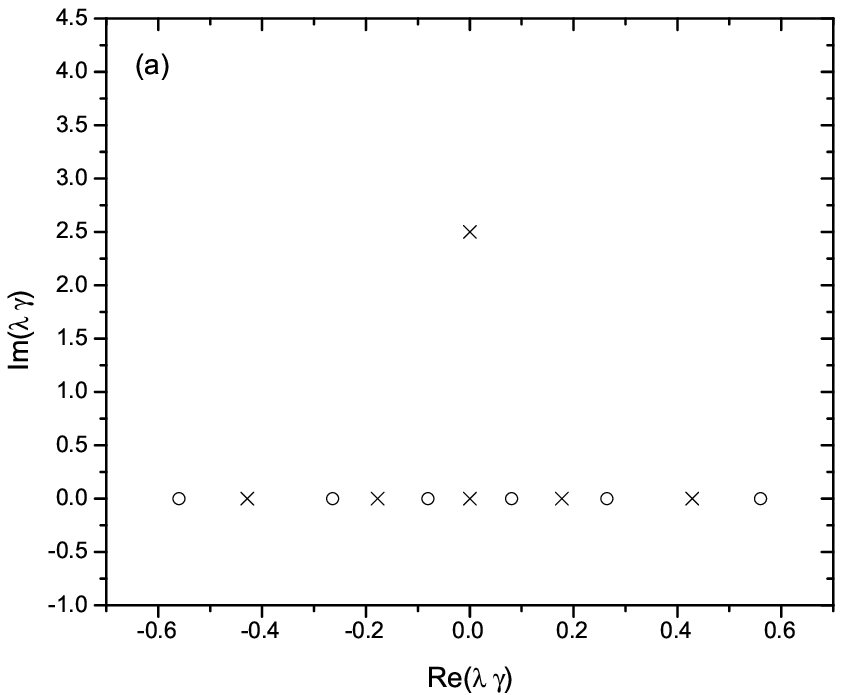}
\end{center}
\end{figure}
\begin{figure}[ht]
\begin{center}
\includegraphics{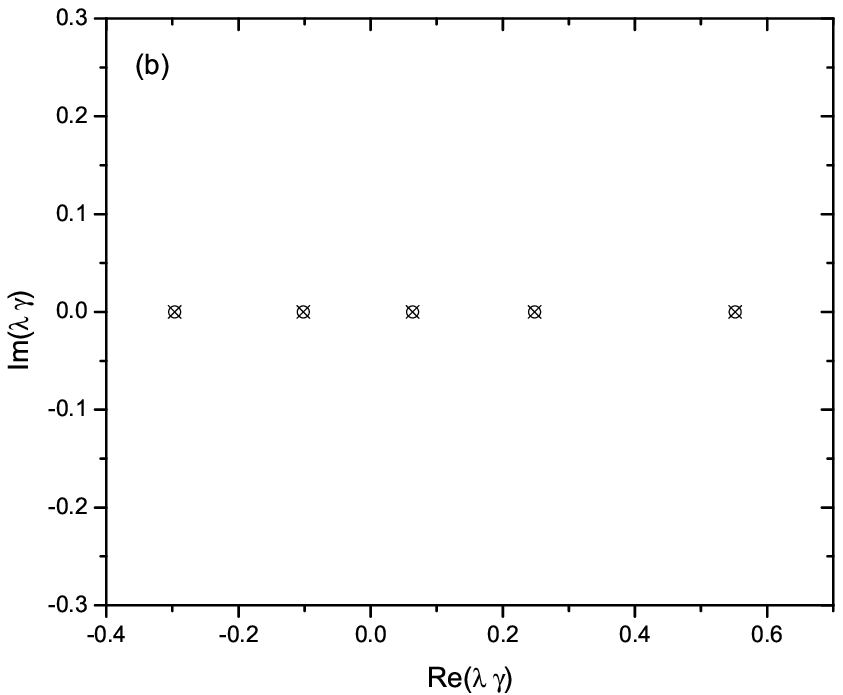}
\end{center}
\end{figure}
\begin{figure}[ht]
\begin{center}
\includegraphics{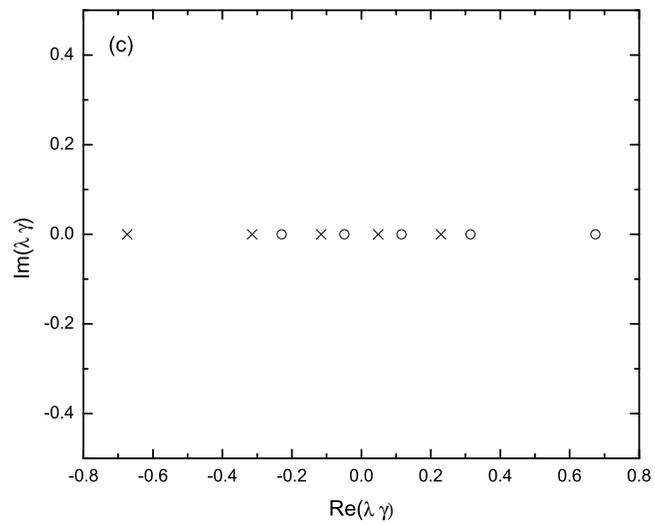}
\end{center}
\caption{\footnotesize{
The Bethe ansatz roots $\lambda_j^{(+)}$ ($\circ$) and
$\lambda_j^{(-)}$ ($\times$) for $\gamma=\frac{\pi}{5}$ and $L=12$. The roots refer
to the dimensions (a) $X_{0,0}^{\frac{1}{2},\frac{1}{2}}(\gamma)$ , (b) $X_{1,1}^{\frac{1}{2},\frac{1}{2}}(\gamma)$
and $X_{1,1}^{\frac{1}{2},-\frac{1}{2}}$.}}
\end{figure}

\end{document}